\documentclass[fleqn,usenatbib]{mnras}

\usepackage{newtxtext,newtxmath}

\usepackage[T1]{fontenc}
\usepackage{multirow}
\usepackage{array}
\usepackage{gensymb}
\newcolumntype{P}[1]{>{\centering\arraybackslash}p{#1}}
\usepackage{threeparttablex,longtable,booktabs} 
\usepackage{graphicx}
\usepackage{caption}   
\usepackage{tablefootnote}

\DeclareRobustCommand{\VAN}[3]{#2}
\let\VANthebibliography\thebibliography
\def\thebibliography{\DeclareRobustCommand{\VAN}[3]{##3}\VANthebibliography}

\usepackage{graphicx}	
\usepackage{amsmath}	

\title[X-ray Variability in M31, M81 and Cen\,A]{Probing Periodic and Aperiodic Variability of X-ray Sources in M31, M81 and Centaurus\,A with {\it Chandra}}

\author[J. Zhang et al.]{
Jiachang Zhang,$^{1,2}$\thanks{E-mail: zhangjiachang@smail.nju.edu.cn}
Zhiyuan Li,$^{1,2}$\thanks{E-mail: lizy@nju.edu.cn}
Ziqian Hua, $^{1,2}$
and Tong Bao. $^{3}$
\\
$^{1}$School of Astronomy and Space Science, Nanjing University, Nanjing 210046, China\\
$^{2}$Key Laboratory of Modern Astronomy and Astrophysics (Nanjing University), Ministry of Education, Nanjing 210046, China\\
$^{3}$ INAF -- Osservatorio Astronomico di Brera, Via E. Bianchi 46, 23807 Merate, Italy
}

\date{Accepted 2025 December 12. Received 2025 December 12; in original form 2025 October 14}
\pubyear{\the\year{}}

\begin{document}
\label{firstpage}
\pagerange{\pageref{firstpage}--\pageref{lastpage}}
\maketitle

\begin{abstract}
Based on archival \textit{Chandra} observations, we present a systematic timing survey of several hundred X-ray sources in M31, M81, and Centaurus~A, mostly low-mass X-ray binaries (LXMBs), focusing on searching and characterizing aperiodic and periodic variability within single observation. We identify flares in 24 sources in M31, 5 in M81, and 26 in Cen~A; several display recurrent events. Flare durations span from tens of seconds to a few $10^{4}$ s, with peak luminosities of $10^{37}$--$10^{40}\ \mathrm{erg\ s^{-1}}$ and low {\rm flare duty cycles} of $4.9\times10^{-6}$--$3.5\times10^{-2}$. 
Dipping events are found in 8 sources in M31, 1 in M81, and 5 in Cen~A, including two repeaters. On multi-epoch baselines, the standard deviation of the source luminosity correlates linearly with the mean luminosity, with a coefficient of $0.49$ (M31), $0.30$ (M81), and $0.67$ (Cen~A), indicating galaxy-to-galaxy diversity. No statistically significant periodic signals are detected in M81 or Cen~A, which, along with several periodic signals previously found among the M31 sources, can be understood considering a joint effect of our detection sensitivity and intrinsic distributions of the orbital period and X-ray luminosity of LMXBs. The ensemble of short-duty-cycle flares, a mix of recurrent and isolated dips, and galaxy-dependent rms--flux factor, supports a picture in which stochastic accretion-rate fluctuations modulate luminosity on $\sim$10--$10^{4}$ s. 
Conducted at known distances and across distinct host environments, this extragalactic survey provides uniform flare/dip samples and rms--flux scalings for bulge-dominated fields, offering empirical constraints for accretion physics and illustrating the promise of timing analyses in external galaxies using the \textit{Chandra} archive.
\end{abstract}




\section{Introduction}

Temporal variability is a key property of X-ray binaries (XRBs), providing direct constraints on accretion physics and the physical properties of compact objects. The variability of XRBs spans timescales from milliseconds to years, including phenomena such as pulsations, orbital modulations, quasi-periodic oscillations, and thermonuclear bursts, as well as longer-term changes driven by variations in the mass-accretion rate \citep{2005MNRAS.359..345U,vanderKlis_2006,2008ApJS..179..360G,2002MNRAS.332..231U}.

In Galactic XRBs, extensive timing studies have revealed both periodic and aperiodic behaviour. Aperiodic variability includes thermonuclear (type~I) X-ray bursts, flares, and stochastic variability in the persistent emission. Systematic efforts, such as the \textit{Multi-Instrument Burst Archive} (MINBAR), have catalogued thousands of type~I bursts, enabling statistical analyses of recurrence, energetics, and correlations with accretion state \citep{2020ApJS..249...32G}. Beyond the classical short bursts, intermediate-duration and ultralong bursts have also been identified, further extending the phenomenology of thermonuclear burning \citep{2020MNRAS.494.2509A,2023MNRAS.521.3608A}. 
Periodic signals, on the other hand, include orbital eclipses and pulsations. Large compilations of Galactic low-mass X-ray binaries (LMXBs) have established robust samples for such periodicity studies, providing catalogues of orbital periods, pulsations, and other coherent signals in LMXBs\citep{2023hxga.book..120B,2023A&A...675A.199A}. These catalogues supply the demographic context necessary for interpreting accretion-driven variability and benchmarking extragalactic searches. 
At longer timescales, variability between observations is often linked to accretion state transitions. Power spectral analyses reveal broad-band noise components and quasi-periodic oscillations, with characteristic frequencies correlated with luminosity and spectral state \citep{2002MNRAS.332..231U}. The discovery of a linear rms–flux relation across XRBs and active galactic nuclei (AGNs) further points to a common variability process driven by propagating mass-accretion rate fluctuations \citep{2005MNRAS.359..345U}. Together, these results demonstrate that timing provides a powerful and effective diagnostic of accretion flows.

In contrast, timing studies of extragalactic XRBs remain limited, due primarily to photon statistics and monitoring constraints. A handful of periodic sources have been reported in M31, including eclipsing candidates identified from deep \textit{Chandra} observations \citep{2024MNRAS.530.2096Z}. Transient burst events have also been observed, such as those reported in the XMM-Newton observations \citep{2005A&A...430L..45P} and the EPIC-pn XMM-Newton outburst detector (EXOD) survey \citep{2020A&A...640A.124P}, and more extreme cases of ultraluminous X-ray bursts detected from compact objects in nearby elliptical galaxies \citep{2016Natur.538..356I}. Recently, quasi-periodic oscillations and low-frequency timing features have been identified in a number of nearby ultraluminous  X-ray sources (ULXs), e.g., the 54 mHz QPO in M82 X-1 \citep{2003ApJ...586L..61S}, the 20 mHz QPO in NGC 5408 X-1 \citep{2007ApJ...660..580S}, the 202.5 mHz QPO in Holmberg IX X-1 \citep{2006ApJ...641L.125D}, and the recurrent dip of 2 days in M51 ULX-7\citep{2021ApJ...909....5H}, highlighting the potential of extragalactic timing studies to probe accretion under extreme conditions. 
Overall, while Galactic XRBs have yielded a mature picture of temporal variability, extending such studies to extragalactic systems remains at an early stage. Systematic timing analyses of external galaxies are both advantageous and essential to assess population-wide variability properties, constrain accretion physics in diverse environments, and establish links between close binaries and their host galaxies. 

Thus motivated, we select three nearby galaxies, M31, M81, and Centaurus~A (Cen~A or NGC 5128), to perform a systematic timing analysis of their XRB populations, in particular the LMXBs.
The choice of this pilot galaxy sample is motivated by (i) proximity, which maximizes photon statistics and enables the detection of moderately luminous XRBs; (ii) contrasting host environments that sample different XRB demographics and X-ray luminosity functions (XLFs); 
and (iii) deep, multi-epoch coverage enabling both inter- and intra-observation timing analysis.

M31 is the nearest large spiral beyond the Milky Way and provides the highest extragalactic source surface density at arcsecond scales, with extensive archival X-ray observations. At a distance of 0.78 Mpc and with modest, relatively uniform line-of-sight extinction, a single \textit{Chandra} pointing covers large fraction of the bulge, enabling simultaneous monitoring of hundreds of XRBs at a time \citep{2002ApJ...578..114K,2007A&A...468...49V,2013A&A...555A..65H,2016MNRAS.461.3443V}, which is impractical for Galactic XRBs owing to distance uncertainties, differential extinction, and severe source confusion along the Galactic plane. 

M81, the next large spiral galaxy at a distance of 3.6 Mpc, is well-suited for XRB population studies,  thanks to its large angular size,
moderate crowding and comparatively low diffuse background, enabling clean point-source photometry \citep{2003ApJS..144..213S}. 
Above $L_{\mathrm X}\gtrsim10^{37}\ \mathrm{erg\ s^{-1}}$, the bulge population is LMXB-dominated, and repeated {\it Chandra} observing epochs recover a stable, galaxy-wide XLF despite strong source-to-source variability \citep{2011ApJ...735...26S}. Together, these factors provide a high-purity, well-sampled bulge LMXB set.

Cen A (NGC 5128), as the nearest (at a distance of 3.8 Mpc) massive elliptical galaxy, offers an early-type, post-merger laboratory that contrasts with spiral bulges. A prominent dust lane and merger signatures with residual star formation (including jet–ISM triggered sites) produce a structured diffuse environment around an active nucleus and jet \citep{2020MNRAS.498.2766W,2016A&A...586A..45S}. It also hosts a structured hot gas halo that produces spatially varying diffuse X-ray emission \citep{2002ApJ...577..114K,2003ApJ...592..129K}. Despite this complexity, early X-ray studies resolve and classify the point-source population throughout the galaxy and the LMXB population is quantified down to $\approx 2\times10^{36}\ \mathrm{erg\ s^{-1}}$ \citep{2001ApJ...560..675K,2006A&A...447...71V}, but also complements M31 and M81 with a substantial number of luminous sources by virtue of its large stellar mass. 

In this work, we conduct a systematic analysis of periodic signals, short-term flares and dips, and long-term variability for the X-ray sources in the three galaxies, utilizing a wealth of \textit{Chandra} observations taken over the past two decades. These observations provide a rare but promising dataset population studies of extragalactic XRBs. In Section~\ref{sec:data}, we describe the preparation of the X-ray data and the use of source catalogs. In Section~\ref{sec:method}, We describe the method and procedure for searching periodic and aperiodic variability. In Section~\ref{sec:result}, we present the results and address the implications and limitations of our results. A brief summary of our study is given in Section~\ref{sec:sum}.

\section{Data}
\label{sec:data}
The inner regions of M31, M81 and Cen A have been extensively observed by {\it Chandra} with its Advanced CCD Imaging Spectrometer (ACIS)\citep{2001IAUS..205...38G,2003ApJS..144..213S,2001ApJ...560..675K}.
We use the non-grating ACIS observations. This ensures a homogeneous instrumental setup with undispersed imaging, a well-behaved PSF, and relatively simple background, and avoids the complications introduced by grating dispersion (e.g. long spectral arms and strongly position-dependent effective area). 
In order to assess potential contamination of cosmic X-ray background (CXB) sources in the direction of the three galaxies, which turn out to be non-negligible (Section~\ref{sec:CXB}), we also perform a uniform timing analysis on the ACIS data from the {\it Chandra} Deep Field-South (CDFS; \citealt{2017ApJS..228....2L}). 

\begin{figure*}
\centering
\includegraphics[width=0.47\linewidth,trim=5mm 6mm 12mm 6mm,clip]{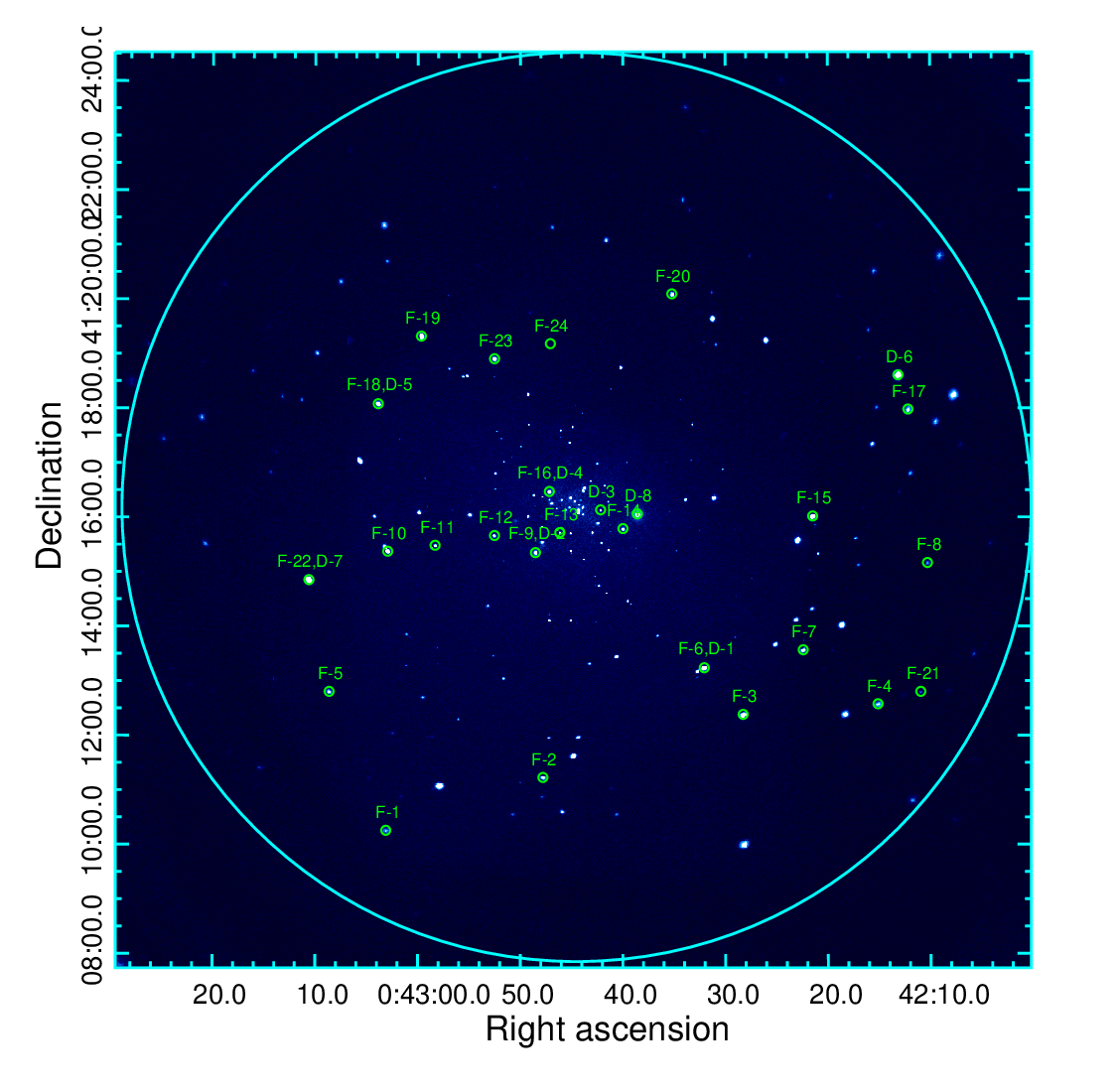}\hfill
\includegraphics[width=0.5\linewidth,trim=7mm 9mm 17mm 10mm,clip]{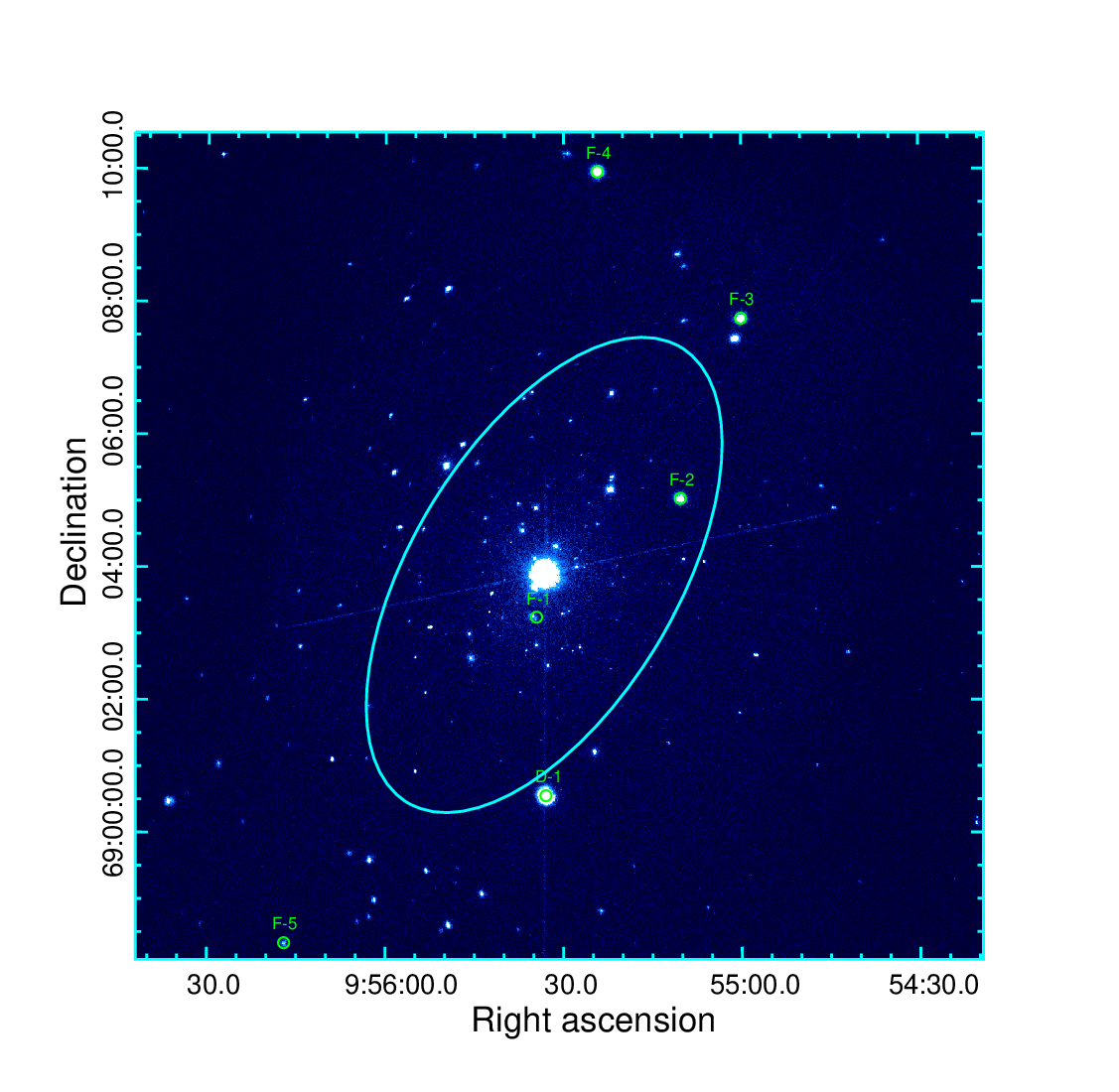}
\vspace{1mm}

\noindent
\includegraphics[width=0.5\linewidth,trim=9mm 9mm 15mm 10mm,clip]{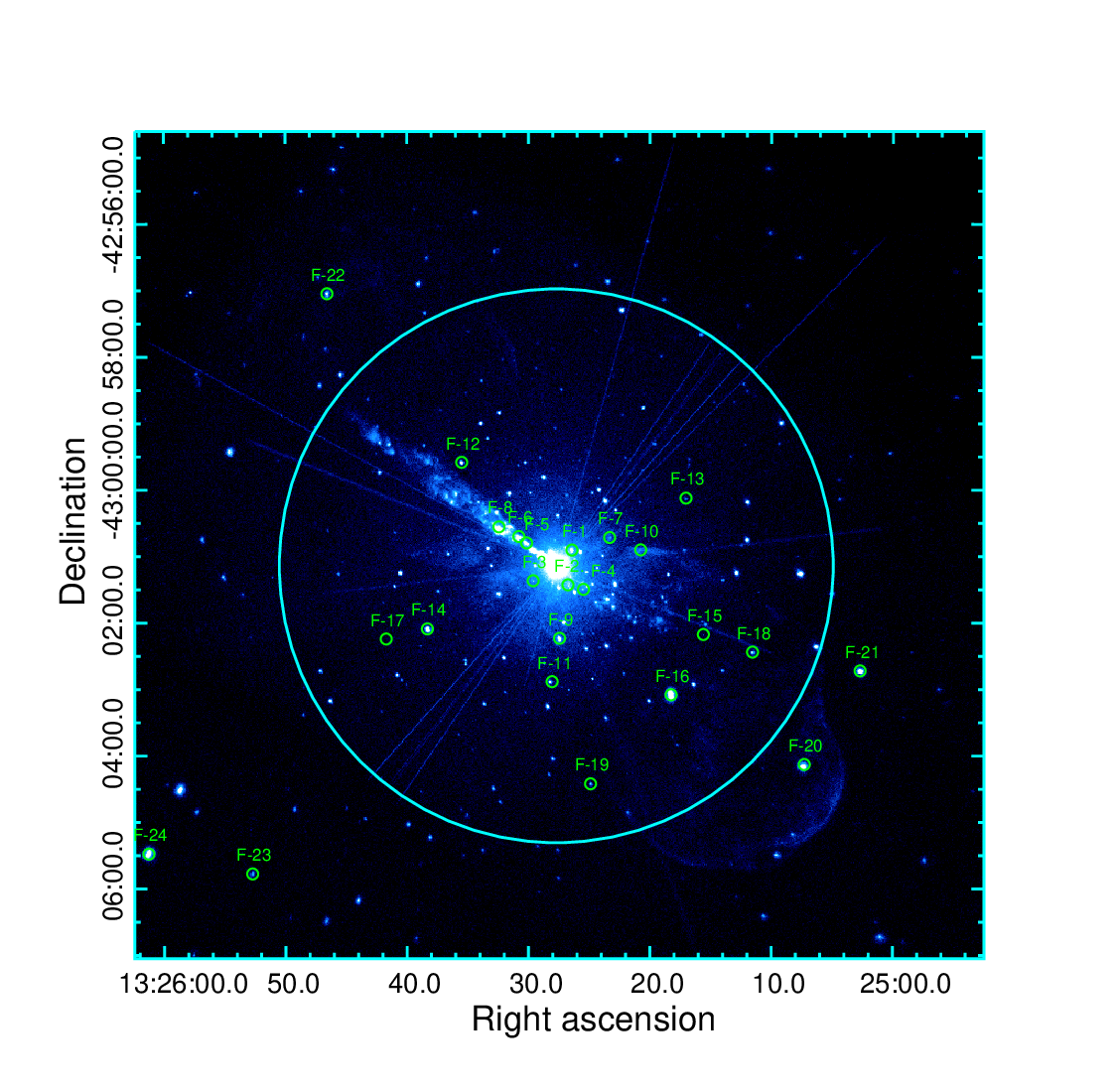}

\caption{Stacked {\it Chandra}/ACIS 0.5--8~keV counts images of M31 (upper left), M81 (upper right), and Centaurus~A (lower panel). X-ray sources showing flares (F) or dips (D) are marked with green circles; labels correspond to the IDs in Tables~\ref{tab:flare} and \ref{tab:dip}. The bulge region used in our analysis is outlined in cyan (see Section~\ref{sec:data}).}
\label{fig:acis}
\end{figure*}



\begin{table*}
    \centering
    \begin{tabular}{ccccccc}
    \hline
    Galaxy  & Distance & Exposure time & Galactic $N_H$ & Bulge radius & Total sources & Bulge sources\\
    \hline
    \hline
M31 & $785\pm25$ kpc & 0.99 Ms 
    & $7.0\times 10^{20}\,\mathrm{cm}^{-2}$ & $500\arcsec$ & 456 & 456 \\
M81 & $3.6\pm0.1$ Mpc & 1.23 Ms 
    & $9.9\times 10^{20}\,\mathrm{cm}^{-2}$ 
    & $4\arcmin\times 2\arcmin$ & 321 & 101 \\
Cen A & $3.8\pm0.1$ Mpc & 1.08 Ms 
      & $2.4\times 10^{20}\,\mathrm{cm}^{-2}$ 
      & $250\arcsec$ & 665 & 312 \\
    \hline
    \end{tabular}
\caption{Basic properties of the three target galaxies. Listed are the adopted distances (with statistical uncertainties), the total \textit{Chandra} ACIS exposure used in this work, the line-of-sight Galactic hydrogen column density $N_{\rm H}$, the definition of the bulge region, and the numbers of X-ray sources in the full field and within the bulge. 
Distances are taken from \citet{2005MNRAS.356..979M} for M31, \citet{2010ApJ...718.1118D} for M81, and \citet{2010PASA...27..457H} for Centaurus~A. 
Galactic $N_H$ values are acquired from the HI 4 Pi Survey \citep{2016A&A...594A.116H}, https://heasarc.gsfc.nasa.gov/cgi-bin/Tools/w3nh/w3nh.pl}
    \label{tab:galaxy}
\end{table*}
The data in each observation were uniformly reprocessed, following the standard procedure in \cite{2018ApJS..235...26Z} and using CIAO v4.14\footnote{https://cxc.cfa.harvard.edu/ciao} and calibration data files v4.9.8 \footnote{https://cxc.cfa.harvard.edu/caldb}. The CIAO tool \texttt{acis\_process\_event} were used to reprocess the level 1 event files to create level 2 event files, which corrected charge transfer inefficiency and filtered cosmic-ray afterglows.  We only included data from the I0, I1, I2 and I3 chips for observations if the aim-point was on I3 and data from the S2 and S3 chips if the aim-point was on S3. Inspection of the integrated light curve of each observation shows that the instrumental background remains quiescent for most intervals. Therefore, we retained all science exposures for M31, M81 and Cen A, which help maximize the temporal baseline of the point sources. For the CDFS,
ObsIDs 1431, 16176, 16184, 17542  were discarded because they were affected by strong particle flares \citep{2017ApJS..228....2L}. 
The resultant total exposure is 0.99, 1.23, 1.08 and 6.92 Ms for M31, M81, Cen A and CDFS, respectively.
A log of the {\it Chandra} observations of M81 and Cen A is given in Table~\ref{tab:obsCenA} and Table~\ref{tab:obsM81}, respectively, while information for  the M31 and CDFS observations can be found in \cite{2024MNRAS.530.2096Z} and \cite{2017ApJS..228....2L}, respectively.

All photon arrival times have been corrected to the Solar system barycenter.
We aligned the astrometry of each ACIS observation by CIAO tool \texttt{reproject\_aspect}. The observation with the longest exposure time (Obs 14196 for M31, Obs 18047 for M81 and Obs 20794 for Cen A) was used as the reference frame for this alignment process. 
The stacked counts map over the energy range of 0.5--8.0 keV is created and shown in Figure~\ref{fig:acis} for the three galaxies. Exposure maps and PSF maps of individual observations were generated, with the PSF maps computed with a specified enclosed count radius (ECR). A fiducial source spectrum was employed for both the exposure and PSF maps.
We use an absorbed power-law model with a photon index of 1.7, representative of extragalactic X-ray binaries \citep{2004ApJ...611..846K,2013A&A...555A..65H}, and a hydrogen column density consistent with the Galactic foreground value, which is $N_{\rm H}=7.0\times 10^{20} \rm cm^{-2}$, $9.9\times10^{20}\ \rm cm^{-2}$,  $2.4\times10^{20}\ \rm cm^{-2}$ for M31, M81 and Cen A, respectively.\footnote{Values are acquired from the HI 4 Pi Survey \citep{2016A&A...594A.116H}. https://heasarc.gsfc.nasa.gov/cgi-bin/Tools/w3nh/w3nh.pl}

For the timing analysis, we started with the {\it Chandra} Source Catalog v2.0\footnote{https://cxc.cfa.harvard.edu/csc/} for M81 and Cen A, the source catalog of M31 as in \cite{2024MNRAS.530.2096Z}, and the source catalog of the CDFS as in \cite{2017ApJS..228....2L}. 
We restricted the source extracting region within a projected galactocentric distance of $500\arcsec$ for M31, $720 \arcsec$ for M81 and $1000 \arcsec$ for Cen A. 
A raw list of 456 sources for M31, 321 sources for M81 and 665 sources for Cen A are included in the timing analysis focusing on the search for periodic signals (Section~\ref{sec:glmethod}) and short-term aperiodic variations (Section~\ref{sec:tsbb}).

In this work we focus on the X-ray binary population associated with the bulges of M31, M81, and Cen~A. The bulge regions of these galaxies are dominated by old, low-mass stellar populations with little or no ongoing star formation \citep[e.g.][]{2004MNRAS.349..146G,2004ApJ...611..846K}, so the short-lived high-mass X-ray binaries are expected to contribute negligibly to the bulge X-ray source population. We therefore assume that the bulge X-ray binaries are predominantly LMXBs and treat them as such in the following analysis. 
We further define a subset of ``bulge sources", for which the \textit{Chandra}/ACIS coverage is most complete and the exposure is relatively homogeneous. For M31, these are the same 456 sources enclosed within the central $500^{\prime\prime}$ (corresponding to a projected physical radius of $\sim$1.9 kpc).
For M81, we follow \citet{2011ApJ...735...26S} to define the bulge region as an elliptical region of $4\arcmin \times 2\arcmin$ (corresponding to $\sim$4.2 kpc $\times$2.1 kpc) with a position angle of $149\degree$, resulting in 101 sources. 
For Cen A, we use a circular region of a galactocentric radius $250\arcsec$ (corresponding to a projected physical radius of $\sim$4.6 kpc), yielding 312 sources.
These bulge sources, most likely LMXBs, are further analyzed for their long-term variability (Section~\ref{sec:km}).

For each source in each observation, we adopted a default 90 percent ECR  to extract the source events, focusing on the energy range of 0.5-8.0 keV. Source events from all observations form the time series of a given source, which are fed to the variability searching algorithm and the period searching algorithm (see detalis below). Local background counts were extracted from a concentric annulus with inner-to-outer radii of 2--4 times the 90 per cent ECR, masking pixels falling within the 90 percent ECR of neighbouring sources, if any. 
The total source counts,
background counts, source area, background area, and local exposure were fed to
the CIAO tool {\it aprates} to calculate the net photon flux
and associated error, which accounts for the Poisson statistics in the
low-count regime.
For a given source in a given observation, if the 3\,$\sigma$ lower limit of the photon flux hits zero, the source is considered non-detected and the 3\,$\sigma$ upper limit of the source photon flux is derived. 
We use the same incident spectrum as for the exposure map to convert the photon flux into the absorbed energy flux, and adopt a distance of 785 kpc for M31, 3.63 Mpc for M81 and 3.81 Mpc for Cen A, to convert the flux into an absorbed luminosity. 

\section{Methods}
\label{sec:method}
\subsection{Intra-observation aperiodic variability: the two-step Bayesian Blocks algorithm}
\label{sec:tsbb}
We search for and characterize short-term variability, in particular flares and dips, in the time series using the Bayesian Blocks algorithm, which finds a globally optimal piecewise constant function representative of the event rate \citep{1998ApJ...504..405S}. 
Specifically, the events of each observation of a given source are partitioned into $K$ contiguous blocks, with block $k$ spanning an live time $T_k$ and containing $N_k$ source counts. 
Each block is enclosed by the so-called {\it change points}, within which the event rate is statistically consistent with a constant. 
For a homogeneous Poisson process with rate $\lambda_k$ on block $k$, the logarithm of block likelihood is

\begin{equation}
\ln \mathcal{L}_k =N_k \ln \lambda_k -\lambda_k T_k ,
\label{eq:bb1}
\end{equation}
which is maximized at $\widehat{\lambda}_k = N_k/T_k$. 
Substituting $\widehat{\lambda}_k$ back into Eq.~\eqref{eq:bb1} and removing the constant term yield the block “fitness”

\begin{equation}
\mathcal{F}_k \equiv \max_{\lambda_k}\ln\mathcal{L}_k
= N_k \bigl[\ln N_k - \ln T_k\bigr] ,
\label{eq:bb2}
\end{equation}
up to an additive constant independent of the partition. With a geometric prior on the number of blocks, the objective to maximize is
\begin{equation}
\mathcal{S} =\sum_{k=1}^{K}\mathcal{F}_k - K\,{\tt ncp\_prior},
\end{equation}
where $\tt ncp\_prior$ is the per–block penalty that controls overfitting. The empirical mapping between a target false–positive probability $p_0$ and the penalty (for event data) is
\begin{equation}
{\tt ncp\_prior} =4 -\ln(73.53\ p_0N^{-0.478}) ,
\label{eq:ncpprior}
\end{equation}
with $N$ the total number of events in the analyzed time series \citep{2013ApJ...764..167S}.%
\footnote{Equation~\eqref{eq:ncpprior} reflects the corrected form noted by the authors and adopted in \texttt{Astropy}.}

We restrict the search of short-term variability to single-observation light curves with more than 50 counts, leaving 4512/27575, 853/6362, and 2671/17815 (i.e. $\approx$16\%, 13\%, and 15\%) light curves for M31, M81, and Cen~A, respectively.
We apply the \texttt{Astropy} implementation (\texttt{astropy.stats.bayesian\_blocks}) with \texttt{fitness='events'}, specifying $p_0=0.05$ to derive ${\tt ncp\_prior}$ via Eq.~\eqref{eq:ncpprior}. 
The Bayesian Blocks algorithm uses a dynamic programming scheme to determine a grid of candidate change points, which optimally partition the photon arrival time series into statistically distinct segments. The count rates change discontinuously at the change points. 
To suppress false positives from background fluctuation, we further apply a post–processing background treatment to the change point candidates from Step~1. We follow the “weighted–events” strategy of \citet{2015A&A...578A..80W}: merge source and background event lists and assign weights $+1$ to source events and $-\alpha$ to background events, with $\alpha$ the ratio of {\it area $\times$ exposure} between the source and background. This yields unbiased change–point timing even under substantial background fluctuations.

A list of change points are thus obtained for each source of interest in the M31, M81, Cen A and CDFS fields. It is noteworthy that these change points naturally include the start and end time of each ACIS observation.

Next, from the segmented light curves, we define a block as {\it flares (dips)} if its count rate is significantly {\it higher (lower)} than the adjacent (left and right) blocks. A quantitative test is applied to define ``significant", according to the binomial probability that the observed counts in a candidate block (with $C$ counts and $T$ duration) could arise from a homogeneous process spanning both the candidate and reference blocks (total counts $N$). The reference block is defined as the longest-duration block in the light curve. The cumulative probability 
\begin{equation}
\label{eq:pc}
    P(X\geq C)=1-F(C-1;N,p)
\end{equation}
is computed using the binomial cumulative distribution function $F$, where the expected success probability $p=T/(T+T_{\rm ref})$. We require $P<2.7\times 10^{-3}$ for a change to be considered statistically significant, corresponding to a $3\sigma$ threshold under Gaussian statistics.
This approach effectively suppresses spurious detections driven by statistical fluctuations, while ensuring sensitivity to genuine short-term variability. 

We further impose a minimal duration cut (i.e., $\gtrsim$ one ACIS frame time of $\sim$3.2\,s). We also  filter out artificial flux variations caused by dither or photon pile-up or sources strongly affected by central AGN. When a source region partly overlaps a bad pixel or CCD gap, dithering can introduce spurious dips with nominal periods of $\sim$1000\,s (yaw) and $\sim$707\,s (pitch). 

\subsection{Long-term, inter-observation variability: the Kaplan-Meier estimator}
\label{sec:km}
To quantify the long-term variability of X-ray sources while properly accounting for non-detections, we employed the Kaplan-Meier (KM) product-limit estimator \citep{1985ApJ...293..192F}. The KM estimator is a non-parametric survival analysis method which reconstructs the cumulative distribution function in the presence of censored data (i.e., upper limits). Specifically, for a sample of $N$ measurements with values $x_i$ and censoring 
indicators $\delta_i$ ($\delta_i=1$ for a detection and $\delta_i=0$ for an upper 
limit), the KM estimator of the survival function is given by
\begin{equation}
    \hat{S}(x) = \prod_{x_j \leq x} 
    \left( 1 - \frac{d_j}{n_j} \right),
\end{equation}
where $d_j$ is the number of uncensored events at $x_j$ and $n_j$ is the number 
of objects ``at risk'' just prior to $x_j$. 
Here, “at risk” means all objects that have neither been detected (i.e. experienced an uncensored event) nor censored at any value smaller than $x_j$; in other words, they could in principle still produce an event at $x_j$ or later. As $x$ increases, objects leave this risk set either when they are detected at some $x_j$ or when they are censored, so $n_j$ decreases monotonically with $x_j$. 
The corresponding cumulative distribution 
is $\hat{F}(x)=1-\hat{S}(x)$. From the reconstructed distribution one can compute 
the statistical moments, such as the mean 
\begin{equation}
    E[X] = \int_0^\infty \hat{S}(x)\,dx,
\end{equation}
and the variance 
\begin{equation}
    {\rm Var}[X] = E[X^2] - (E[X])^2,
\end{equation}
thereby incorporating both detections and upper limits in a statistically 
unbiased manner.

In practice, for each source we extracted the net photon flux or the 3\,$\sigma$ upper limit when the source was not significantly detected with CIAO tool \texttt{aprates}. 
Detected points were entered into the KM estimator as uncensored values, while non-detections were included as censored entries 
The resulting KM-based variance thus incorporates both detections and upper limits, and provides an unbiased measure of the long-term variability amplitude for each source. 
Following the prescription of \cite{2025ApJ...992...82H}, we reconstructed the distribution of fluxes for each source, and then computed the first and second moments to obtain the mean flux $E[X]$ and the intrinsic variance ${\rm Var}[X]$. The square root of the variance gives the standard deviation $\sigma[X]=\sqrt{{\rm Var}[X]}$, i.e. the rms variability amplitude. 
We restricted the long-term variability analysis to sources that are detected in at least five observations and are located within the bulge region as defined in Section~\ref{sec:data}. This requirement ensures that each source has a sufficiently well-sampled light curve for a statistically meaningful characterization of long-term variability and for a robust Kaplan--Meier estimate; sources with only one or two detections (and many upper limits) provide very little leverage on variability and can make the Kaplan--Meier estimator unstable. 
For the M81 sources, we further restricted the analysis on the ACIS-S observations, which have their aim-point close to the galactic center, ensuring a full coverage of most bulge sources and an optical point-spread function that minimizes potential contamination by the X-ray-bright nucleus \citep{2021NatAs...5..928S}.

\subsection{Periodic variability: the Gregory-Loredo algorithm}
\label{sec:glmethod}
The Gregory–Loredo (GL) algorithm \citep{1992ApJ...398..146G} is a Bayesian phase–folding algorithm designed to detect periodic signals in event data. Owing to its robustness against irregular sampling and its effectiveness for data sets with a moderate number of photon events, the GL algorithm has been widely applied in X-ray astronomy. In particular, it has been successfully employed to identify periodic X-ray sources in the Galactic bulge \citep{2020MNRAS.498.3513B}, a sample of Galactic globular clusters \citep{2023MNRAS.521.4257B,2024MNRAS.527.7173B}, the CDFS \citep{2022MNRAS.509.3504B}, and the bulge of M31 \citep{2024MNRAS.530.2096Z}, all based on deep {\it Chandra} observations. 

We adopt the GL algorithm to search for potential periodic X-ray signals in M81 and Cen A, following the procedures detailed in \cite{2024MNRAS.530.2096Z}. We focus on period ranges of (100, 500), (500, 1000), (1000, 5000), (5000, 10000), (10000, 20000) seconds, with a frequency resolution of $10^{-6}$, $2\times10^{-7}$, $10^{-7}$, $5\times 10^{-8}$, $10^{-8}$ Hz, respectively. 
The choice of the lower bound is primarily constrained by the counting statistics, due to the limited count rates of the X-ray sources analyzed. The upper bound, on the other hand, is limited by the longest exposure time available. In order to reliably detect a periodic signal, it is generally required that a single observation contains more than one full cycle of the period of interest. For most of our observations, this condition sets an effective upper limit of $\sim2\times10^{4}$~s.

We run the GL algorithm on both the combined time series and the time series from individual observations. We adopt a probability threshold $P_{\rm GL}$ of 90\% to select the tentative periods returned by the GL algorithm. 
We further filter spurious signals arising from one of the following reasons:

(i) Spurious signals caused by the dither pattern of the ACIS detectors. The nominal periods of ACIS are 1000 s in yaw and 707 s in pitch. When part of the source region falls on to a bad pixel or goes outside the CCD edges, dithering will make a spurious periodic signal. We exclude any signals detected at the dither periods or their harmonics.

(ii) Spurious signals produced by strong aperiodic variability, which, when accidentally occupies certain phase bins, can mimic a periodic signal \cite{2020MNRAS.498.3513B}. We examine the reality of such signals by excluding the part of the light curve affected by strong variability.

\subsection{Counting for interlopers: foreground stars and background AGNs}
\label{sec:CXB}
Given the long effective {\it Chandra} exposures, some X-ray-emitting foreground stars should be present in all three target galaxies. To remove them and obtain a clean extragalactic sample, we cross-matched our source catalogs with the {\it Gaia} catalog DR3 \citep{2023A&A...674A...1G} using a matching radius of $1\arcsec$. Sources with parallaxes $\varpi > 0.1$ mas and significance $\varpi/\sigma_\varpi > 3$ were classified as Galactic foreground objects and excluded from further analysis. In total, we identified 15, 5, and 26 foreground X-ray sources toward M31, M81, and Cen A, respectively, among which 15, 0, 6 are located within the defined bulge region. According to \cite{2017ApJS..228....2L}, there are 12 foreground stars in CDFS catalog. 
These foreground sources are excluded from the statistics of the {\it bona fide} LMXB population of the three hosts. 
Nevertheless, for completeness, we have searched for periodic and aperodic variability in these foreground sources and provided the results in Appendix~\ref{ap:star};  however, no statistically significant periodic signal was detected.

Given the long effective exposure times accumulated in the three galaxies, a non-negligible number of background AGNs is also expected. Although individual X-ray sources cannot be unambiguously classified as background AGNs only based on the X-ray data, their collective contribution behind each galaxy field can be determined statistically. To this end, we first computed the sensitivity maps of the merged \textit{Chandra}/ACIS observations. The sensitivity maps were generated following the method of \citet{Kashyap2010}, which determines the local limiting flux at each pixel by considering the background counts in an annulus, the exposure map, and the PSF size at that position. We adopted a false detection probability threshold of $10^{-6}$ and an 90\% ECR, resulting in pixel-wise limiting sensitivity maps in the 0.5–8 keV band.

The sensitivity maps were then converted into sky coverage curves, i.e., the cumulative survey area as a function of flux limit. Folding these curves with the empirical $\log N$–$\log S$ relations of AGNs \citep{Kim2007}, we estimated the number of background AGNs detectable in each field. Within the bulge regions defined for our study, the expected background AGN numbers are $\sim104.1$ in M31 (within $500\arcsec$), $\sim16.8$ in M81 (within an ellipse of $4\arcmin \times 2\arcmin$, PA=$149^\circ$), and $\sim40.7$ in Cen A (within $250\arcsec$). These values demonstrate that background AGNs constitute a non-negligible fraction (13\%--23\%) of the detected X-ray sources in the direction of the three galaxies. 

We note in passing that accreting white dwarfs (WDs) is not the main concern of the X-ray sources in M81 and Cen A, because their distance is further and  detection limit is higher, and no accreting WD systems will have $L_X>4\times 10^{36} \rm~erg~s^{-1}$(see Figure~\ref{fig:km}). 
In M31, there are \(\sim80\) optical novae with supersoft X-ray counterparts identified in the bulge \citep{2014A&A...563A...2H}. Several periodic or pulsation signals from novae/supernova soft sources have also been reported \citep{2024MNRAS.530.2096Z}. In our observations—restricted to individual exposures of \(\lesssim100\) ks, flare-like variability in WD-accreting systems is infrequent, of modest amplitude and concentrated in the very low energy band \citep{2013A&A...549A.120H,2015A&A...578A..39N}. Accordingly, we do not expect accreting WD systems to affect our results.

\section{Results and discussions}
\label{sec:result}
\subsection{Flares and dips}
\label{sec:flare}
\begin{figure*}
\includegraphics[width=.49\textwidth]{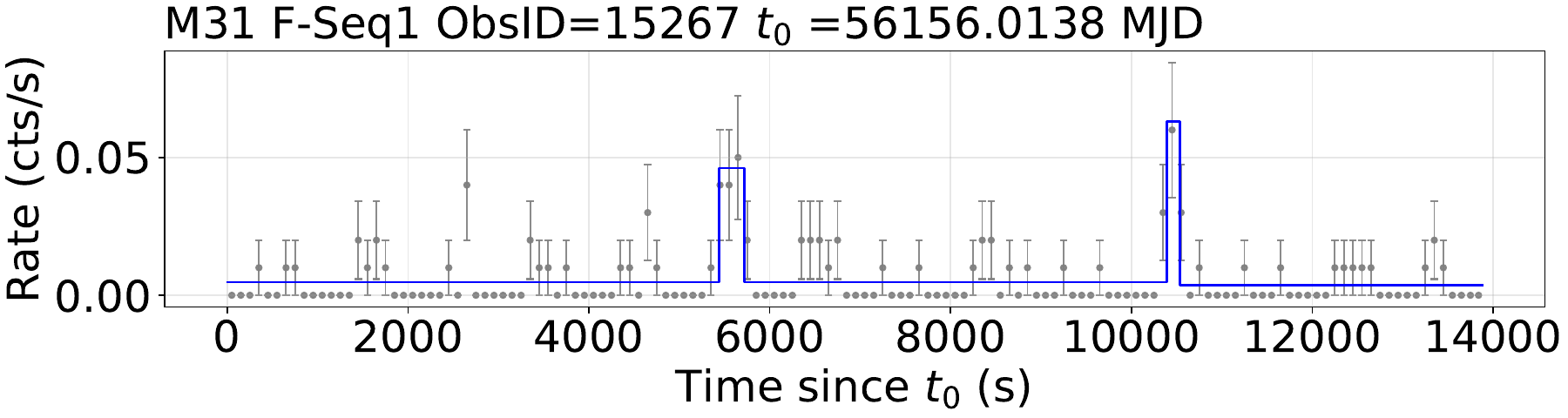}
\includegraphics[width=.49\textwidth]{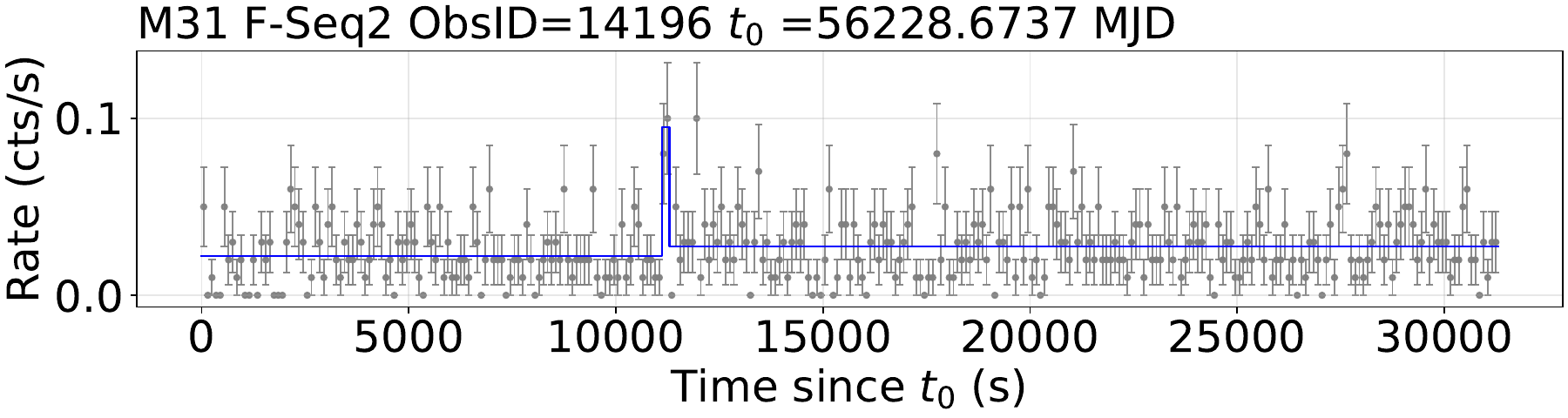}
\includegraphics[width=.49\textwidth]{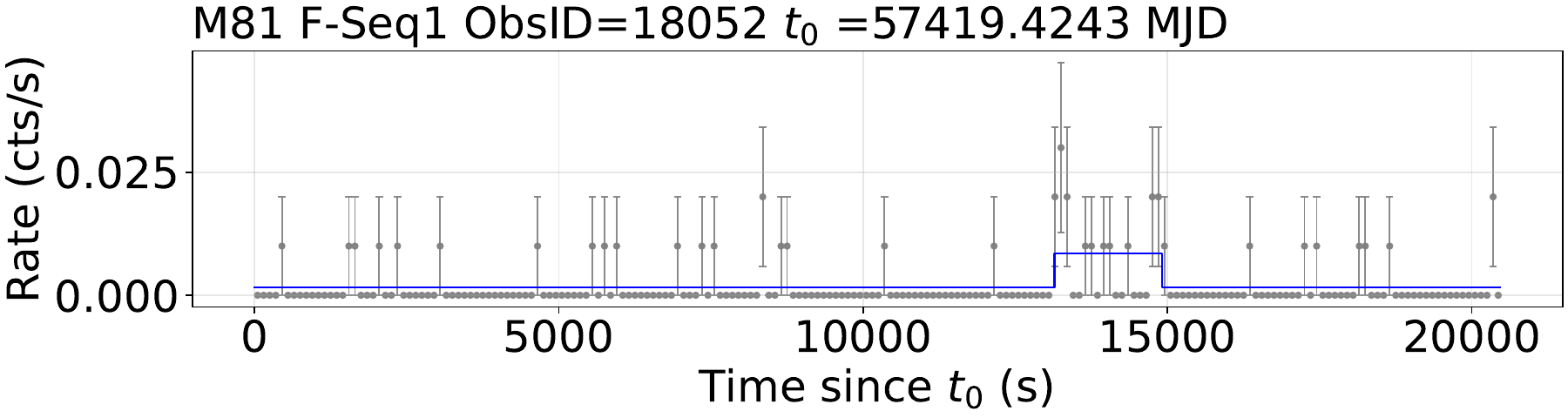}
\includegraphics[width=.49\textwidth]{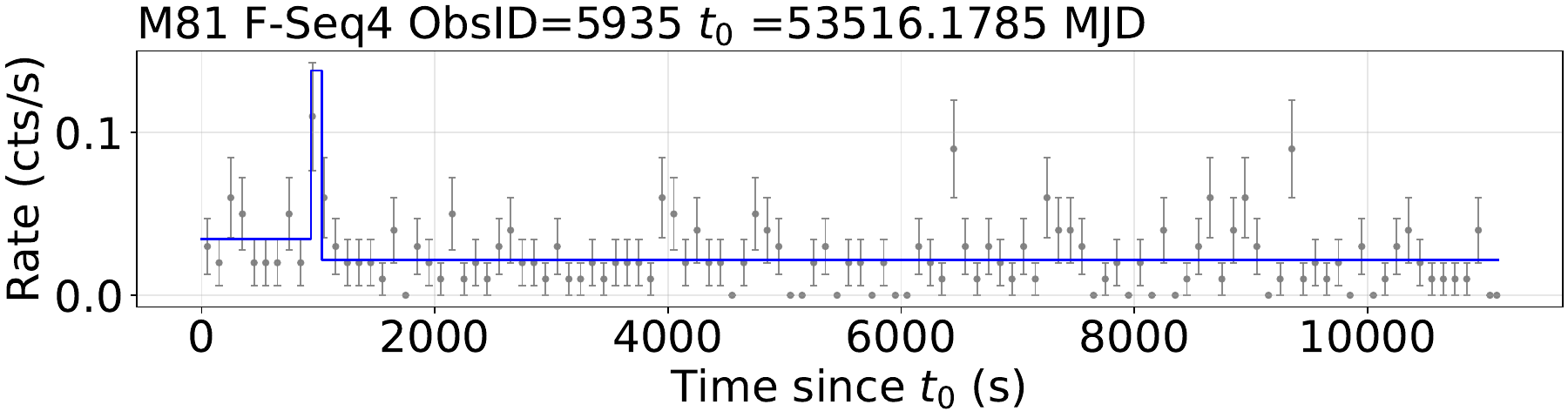}
\includegraphics[width=.49\textwidth]{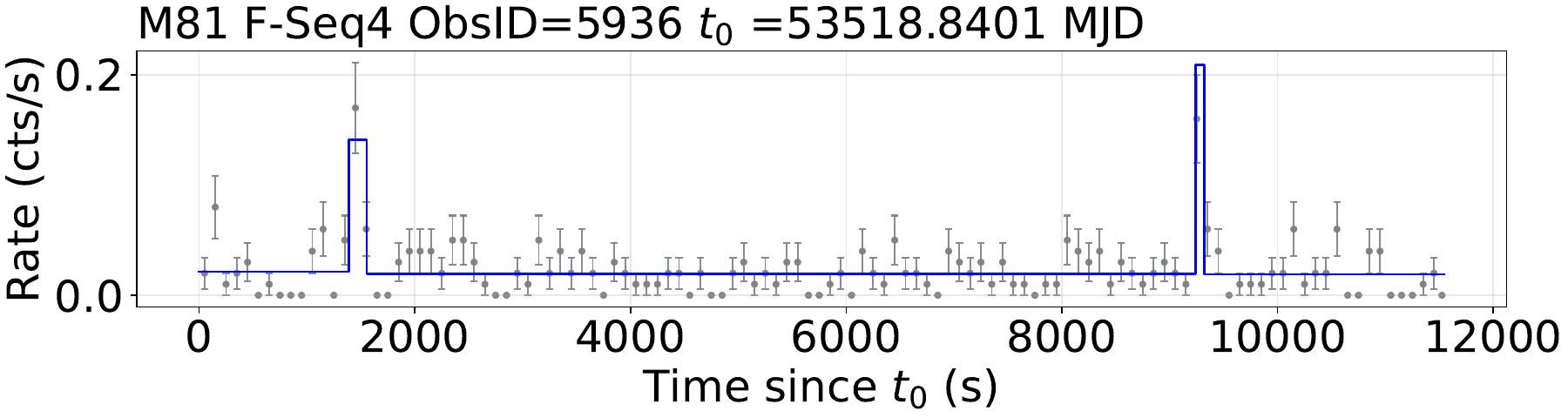}
\includegraphics[width=.49\textwidth]{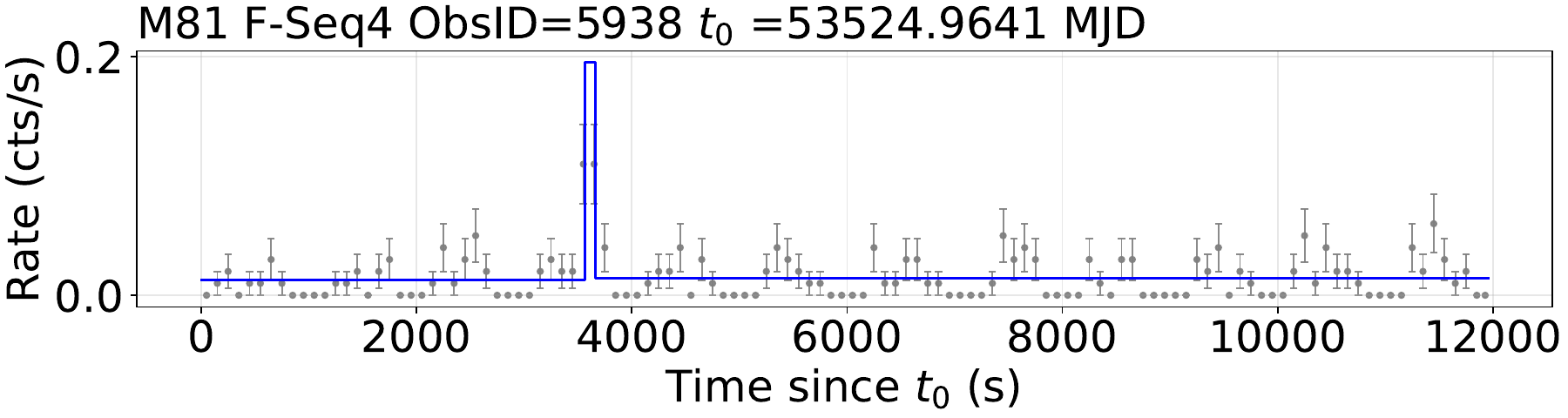}
\includegraphics[width=.49\textwidth]{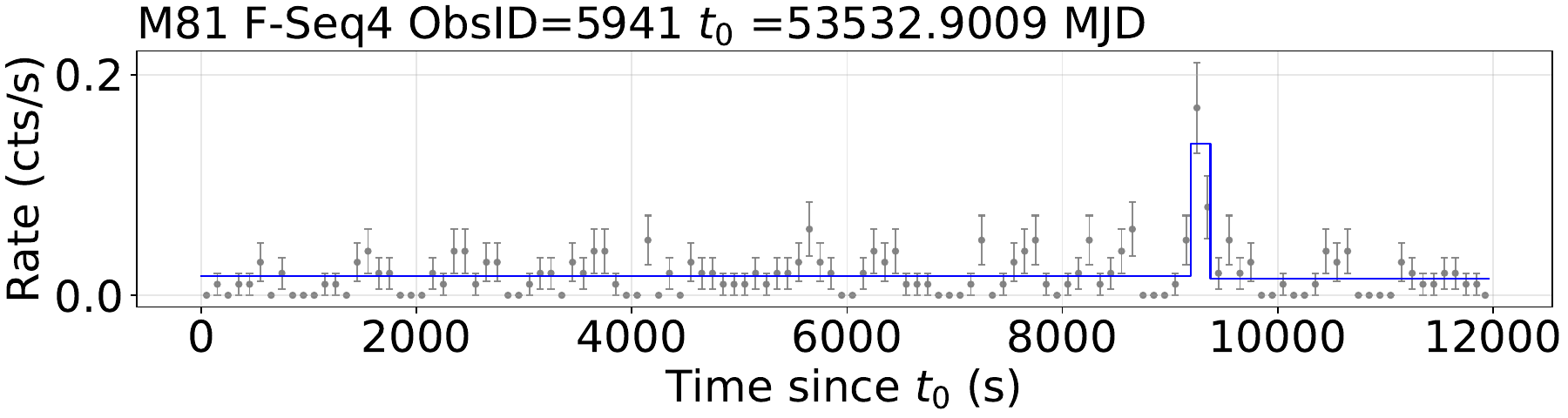}
\includegraphics[width=.49\textwidth]{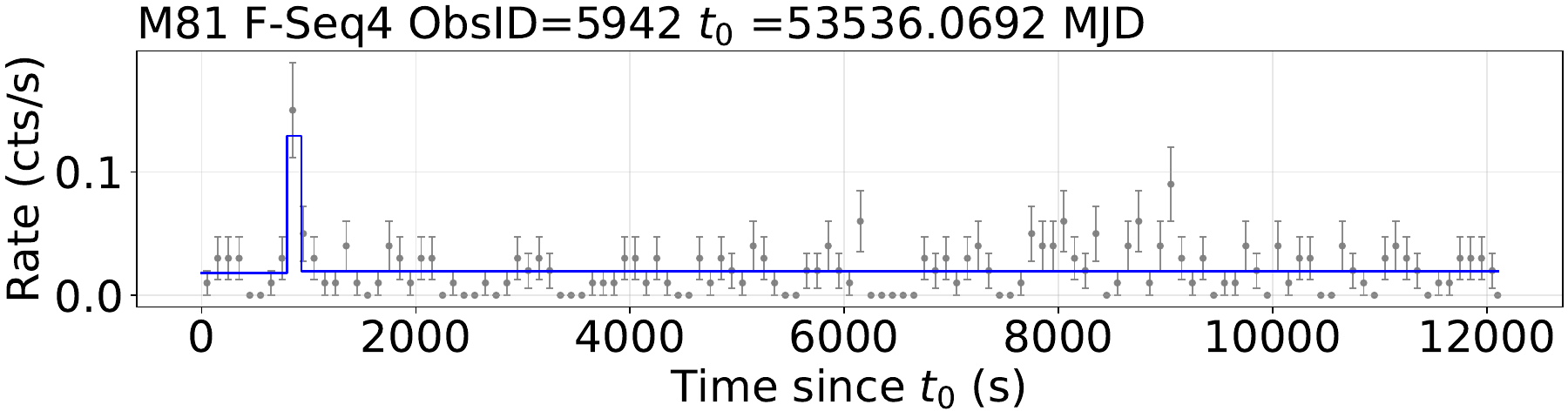}
\includegraphics[width=.49\textwidth]{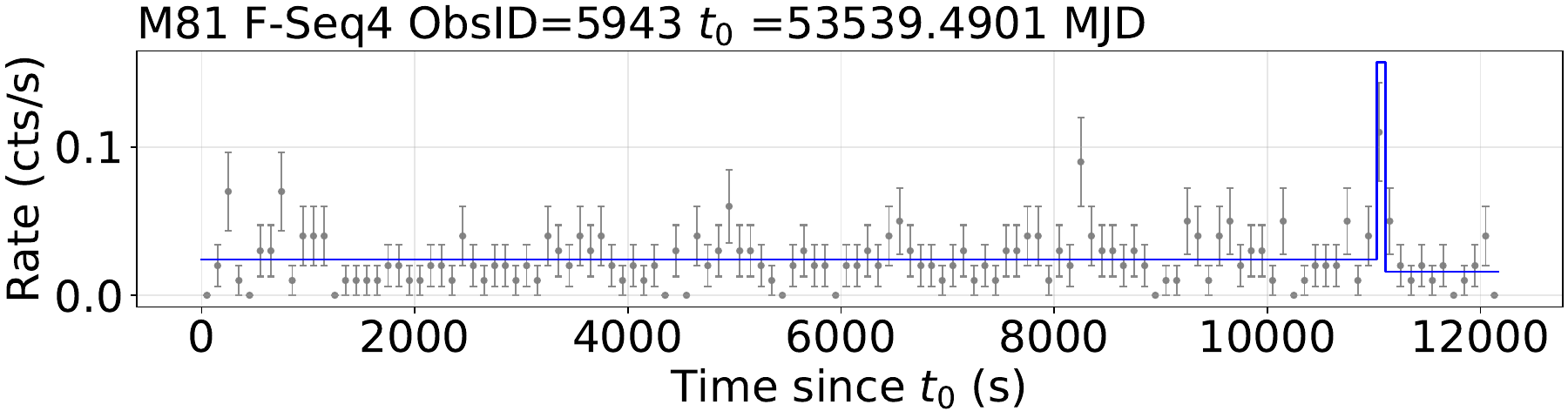}
\includegraphics[width=.49\textwidth]{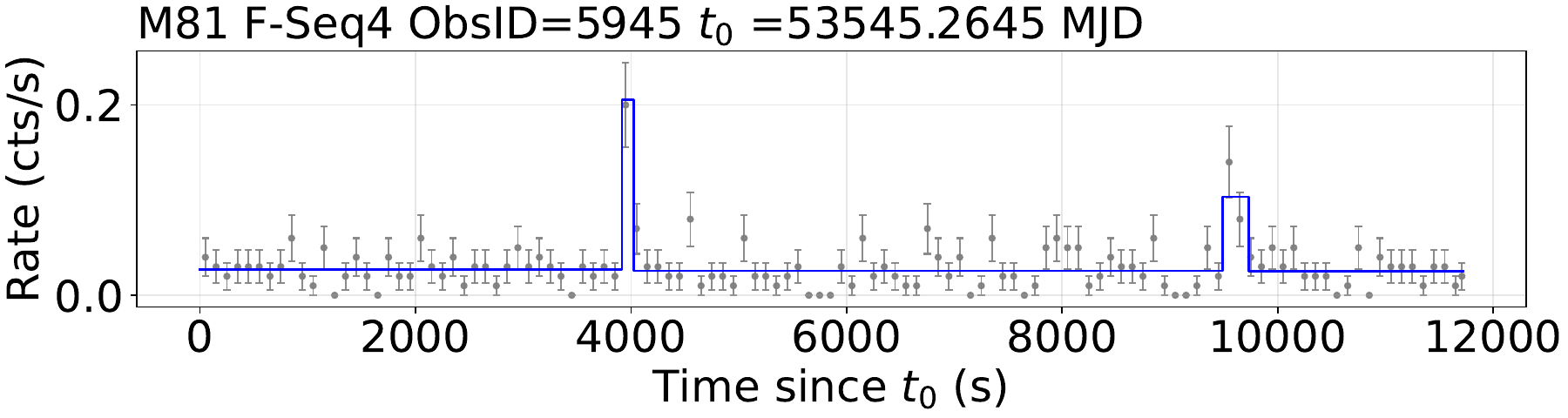}
\includegraphics[width=.49\textwidth]{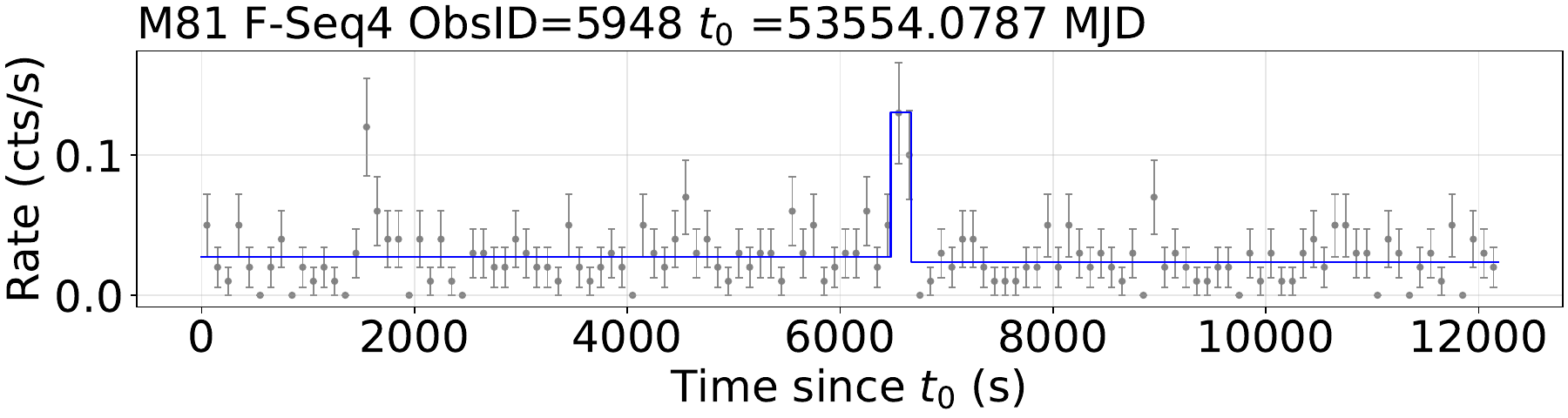}
\includegraphics[width=.49\textwidth]{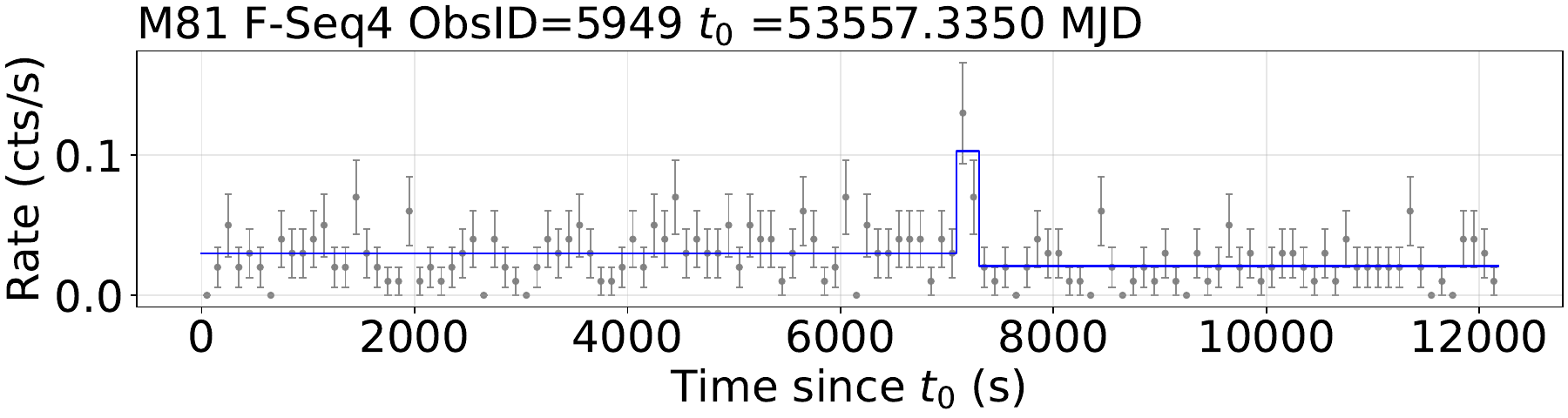}
\includegraphics[width=.49\textwidth]{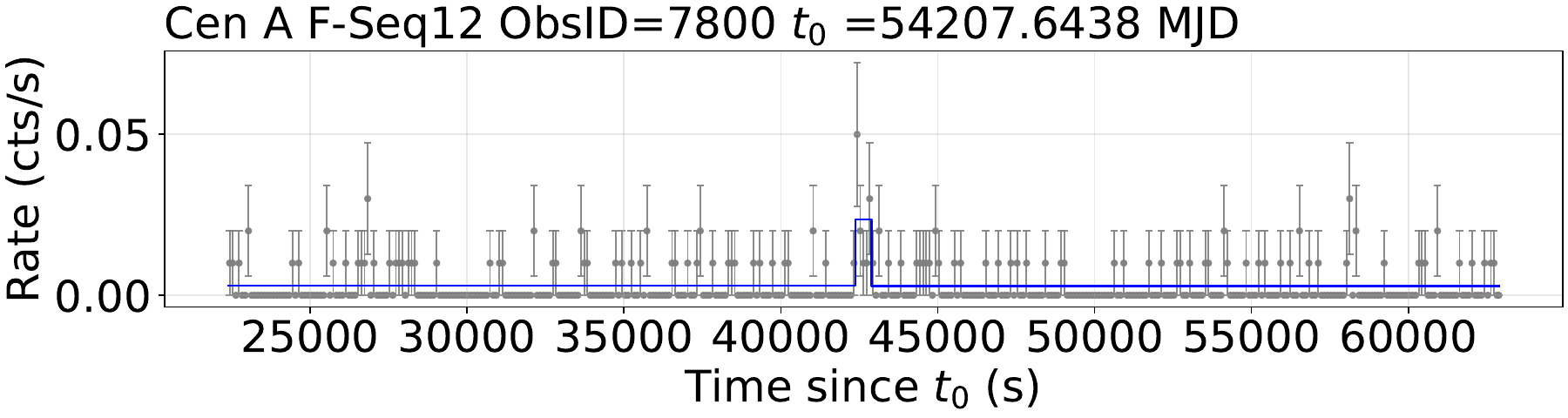}
\includegraphics[width=.49\textwidth]{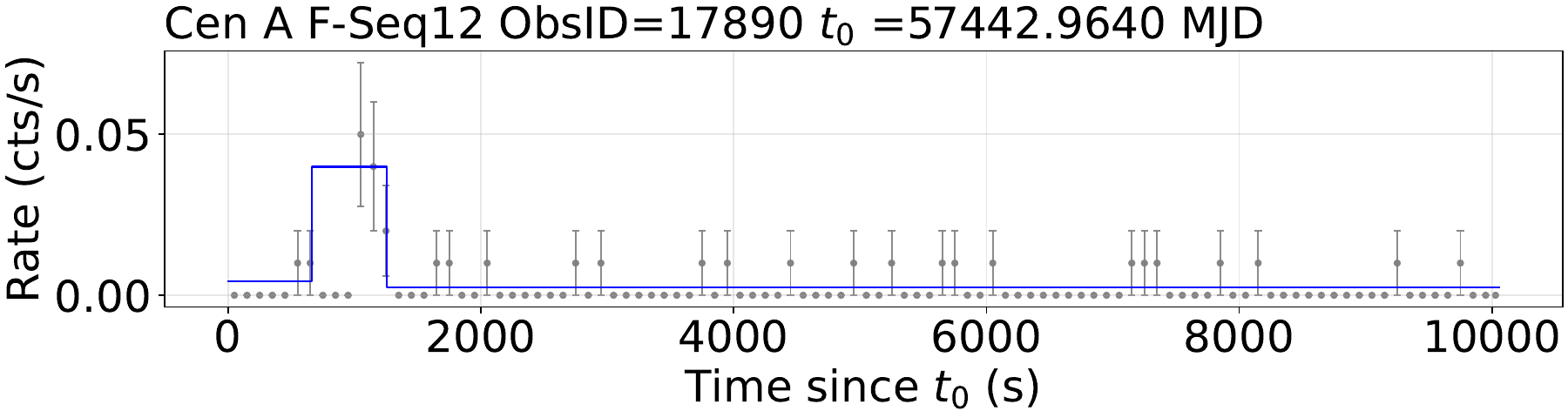}
\includegraphics[width=.49\textwidth]{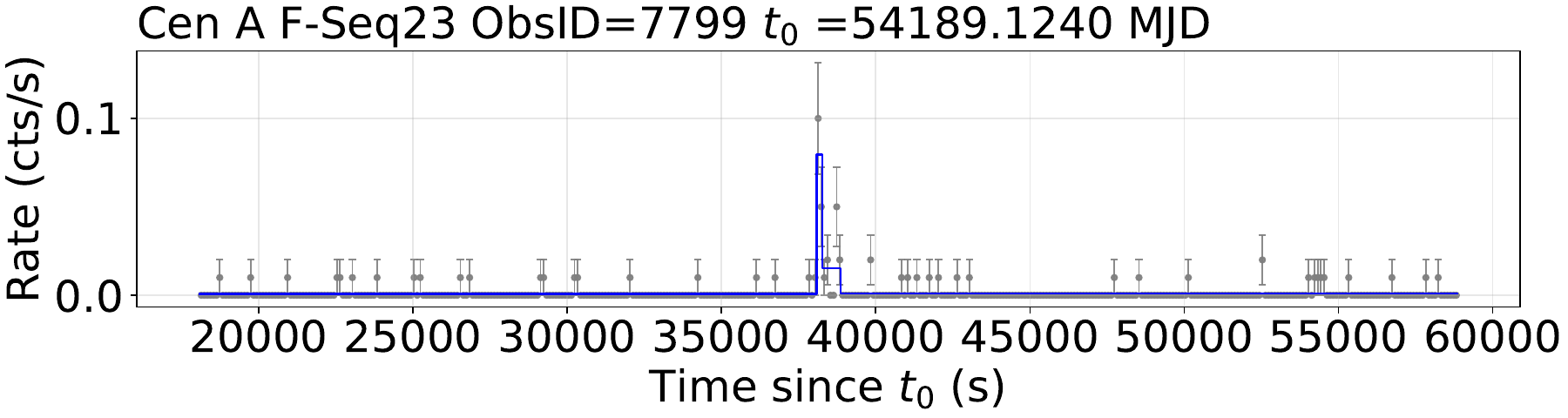}
\includegraphics[width=.49\textwidth]{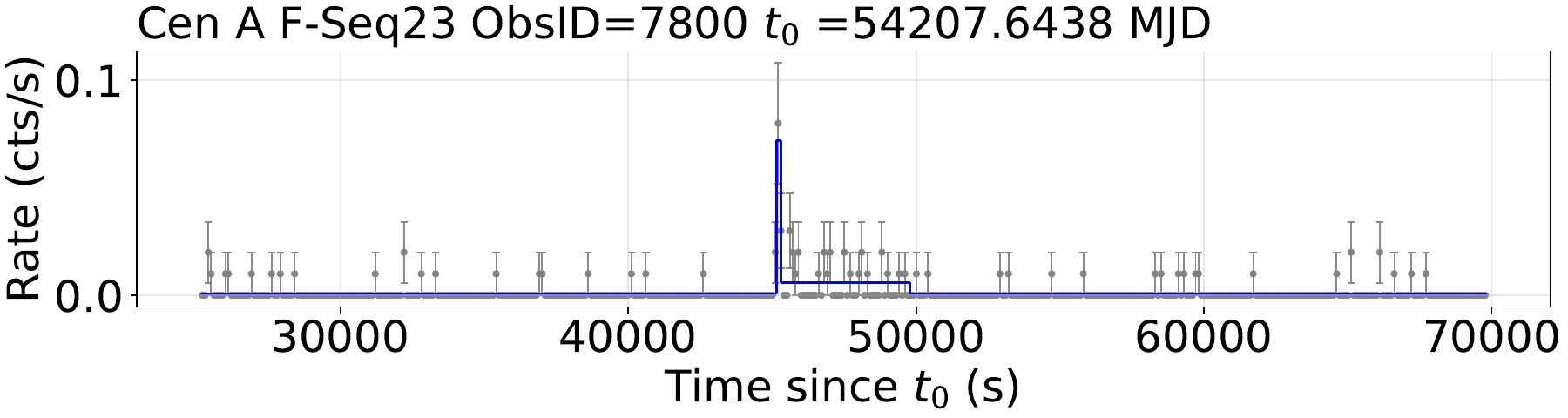}
\includegraphics[width=.49\textwidth]{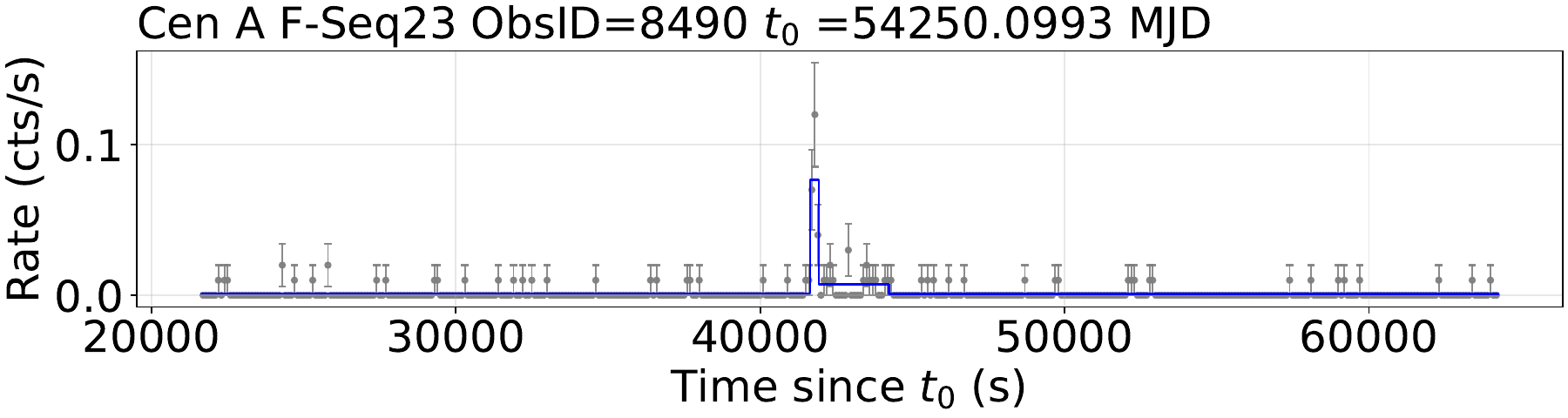}
\includegraphics[width=.49\textwidth]{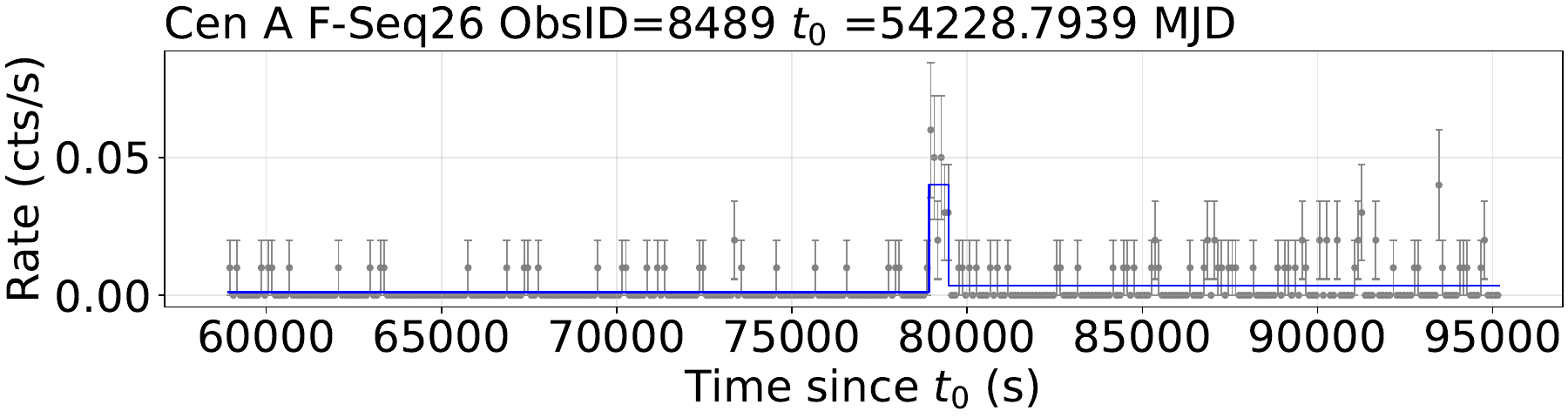}
  \caption{Examples of flares. The original light curve is shown as black data points with error bars. The Bayesian Blocks are shown as blue solid lines. The source information is provided at the upper left corner of each panel.}
  \label{fig:flare}
\end{figure*}

\begin{figure*}
\includegraphics[width=.49\textwidth]{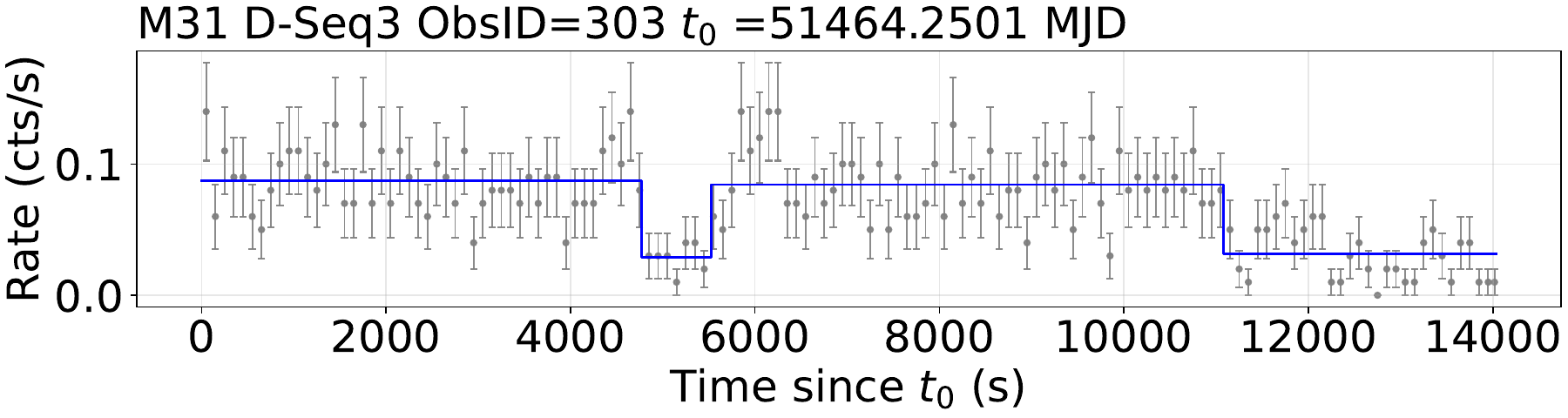}
\includegraphics[width=.49\textwidth]{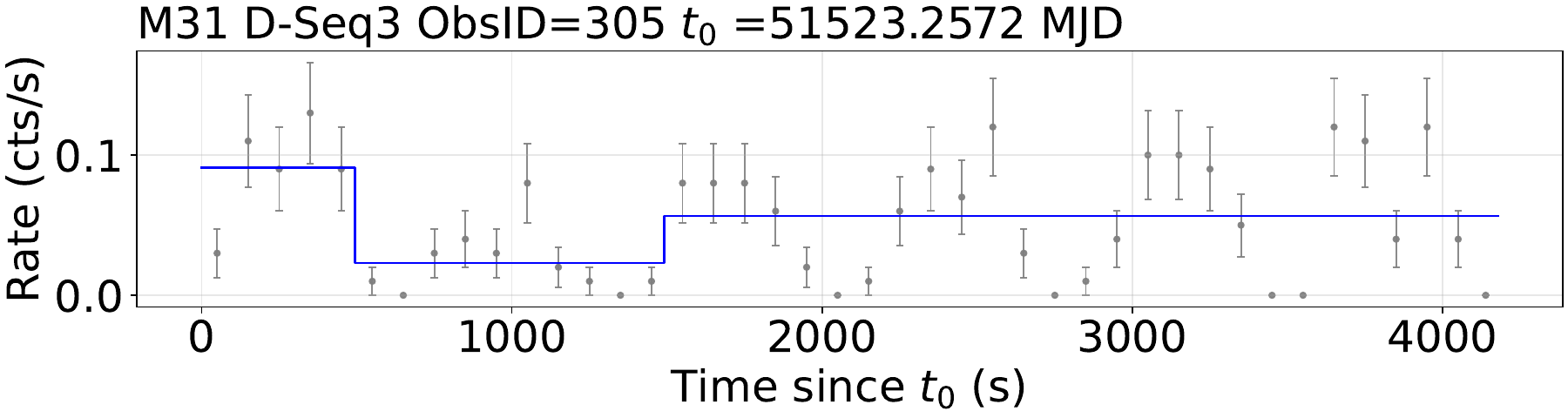}
\includegraphics[width=.49\textwidth]{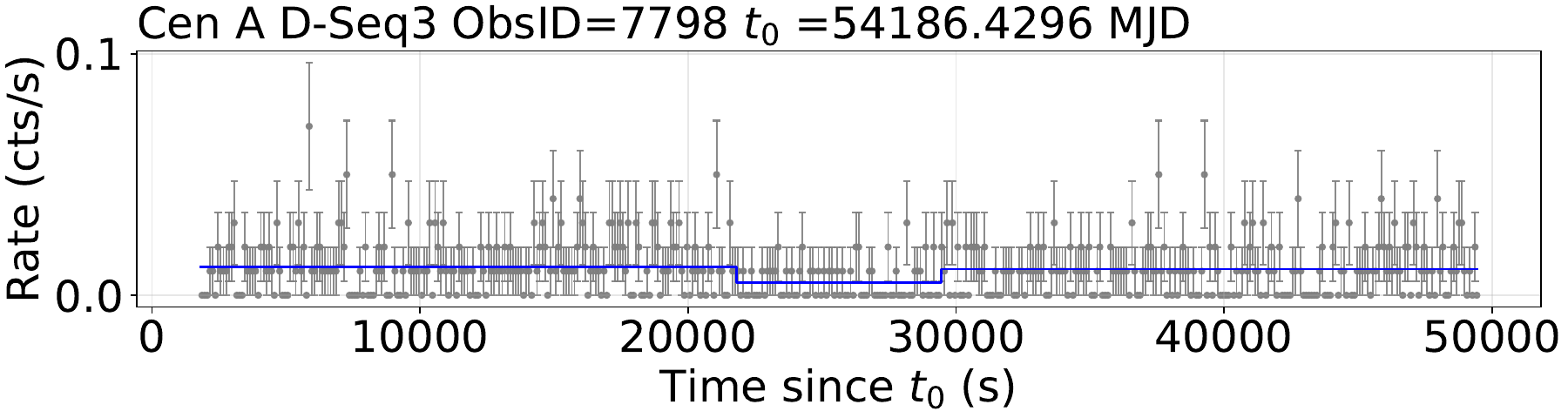}
\includegraphics[width=.49\textwidth]{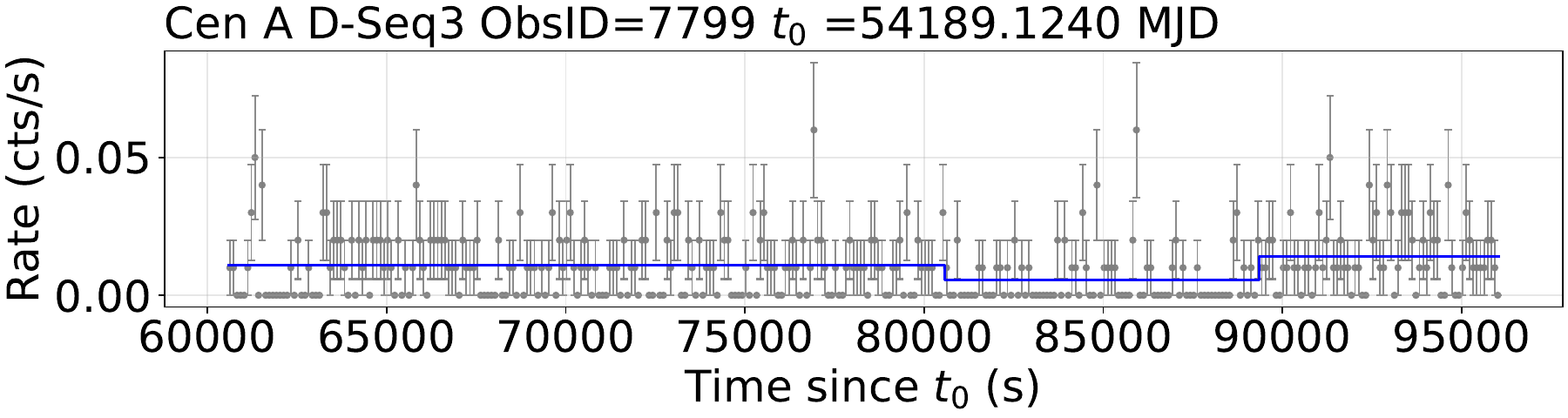}
\includegraphics[width=.49\textwidth]{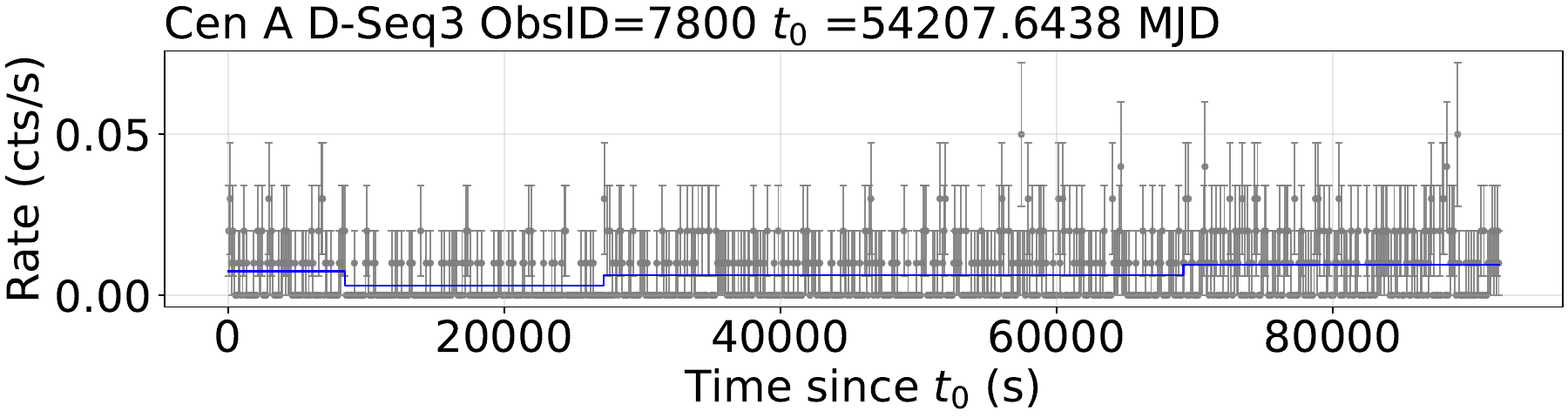}
\includegraphics[width=.49\textwidth]{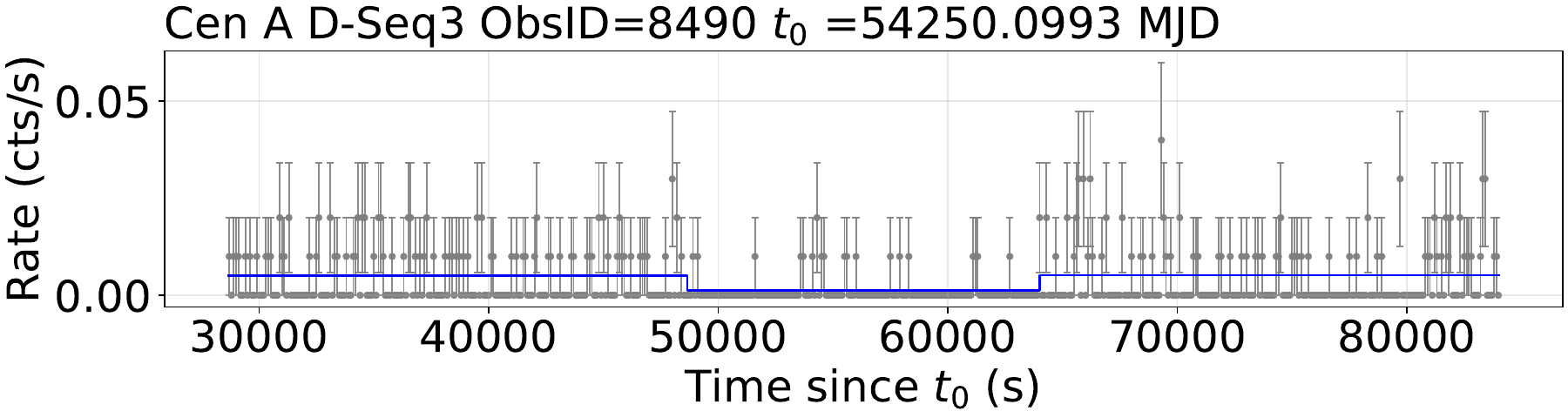}
\includegraphics[width=.49\textwidth]{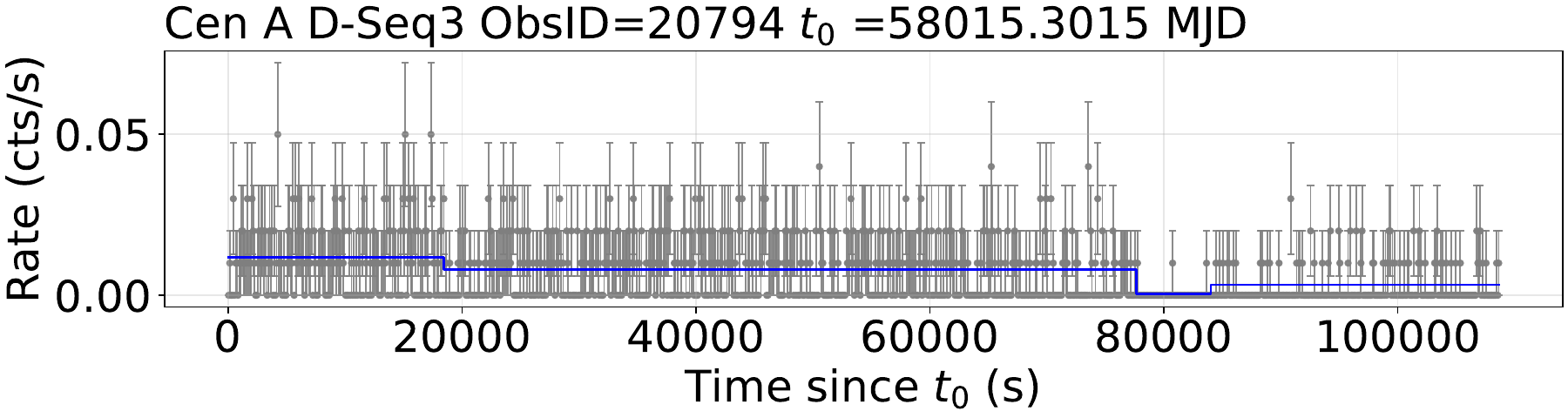}
\caption{Examples of dips. The original light curve is shown as black data points with error bars. The Bayesian Blocks are shown as blue solid lines. The source information is provided at the upper left corner of each panel.}
\label{fig:dips}
\end{figure*}

Following the procedures described in Section~\ref{sec:tsbb}, we detected both flares and dips, with flares being more common. Specifically, we detected flares in 23, 5, and 26 independent sources in M31, M81 and Cen A, respectively, in which several sources show repeated flares. On the other hand, we detected dips in 8, 1, and 5 independent sources in M31, M81 and Cen A. One source in M31 and one source in Cen A show repeated dips. Detailed properties of individual events, including source position, duration, count rate, and luminosity, are listed in Table~\ref{tab:flare} and Table~\ref{tab:dip}. Uncertainties on the count rates, luminosities, and energies are quoted at the $1\sigma$ level. Examples of flares and dips are shown in Figure~\ref{fig:flare} and Figure~\ref{fig:dips}, respectively.

The duration of each flare or dip is defined as the temporal extent of the corresponding block identified by the Bayesian Block algorithm \citep{2013ApJ...764..167S}. 
The flare/dip luminosity is calculated from the source count rate, 
$L = 4\pi d^{2}{\eta}CR$,
where $CR$ is the count rate of the block, $\eta$ is the count rate-to-flux conversion factor (Section~\ref{sec:data}), and $d$ is the source distance.
For the flares, we further subdivide the associated block into five equal temporal bins and adopt the maximum count rate (and corresponding absorbed luminosity) among the five bins as the peak value, which minimizes statistical fluctuations while preserving temporal resolution.

We deliberately adopt a fixed number of equal-width bins, rather than re-applying the Bayesian--blocks method with a different threshold, because the photon statistics of extragalactic X-ray binaries are often low and a more aggressive segmentation would tend to overfit Poisson noise and yield an inhomogeneous number of sub-blocks that depends sensitively on the choice of prior. This simple binning scheme provides a uniform and statistically stable way to characterise the flare peak for all events, without attempting a detailed classification of flare profiles. In this work we only use the brightest bin as a peak indicator and do not attempt to classify the detailed flare morphology (e.g. FRED-like versus symmetric profiles), given the limited number of photons in individual flares.

A number of flares and dips are found in the three galaxies. most likely associated with the LMXB population therein. We provide their basic information, including sky coordinates,
ObsID, flare/dip start time and duration, the peak count rate of the flare, the count rate of the reference block, and the flare energy and peak luminosity. The basic properties of the detected flares are summarized in Table~\ref{tab:flare}, which also has a machine-readable version available online.

For each detected flare, we also derived several characteristic parameters. 
The parameter $\gamma = CR_{\rm ref}/CR_{\rm peak}$ quantifies the ratio of persistent count rate to peak flare count rate. 
The flare rate is defined as $r_{\rm flare}=N_{\rm flare}/t_{\rm total}$, presenting the flare rate measured for each source. 
The intrinsic flare duty cycle is defined as the fraction of each flare cycle occupied by the flare itself:
${\rm DC_{flare}} = \Delta t_{\rm flare}/t_{\rm rec}$, where $t_{\rm rec}=1/r_{\rm rec}$. 
This definition reflects the physical activity level of the source within a typical flare cycle, rather than an observational or instrumental duty cycle.

The dip depth is defined as the fractional decrement relative to a reference block,
depth $= 1 -\frac{CR_{\mathrm{dip}}}{CR_{\mathrm{ref}}}$
where $CR_{\mathrm{dip}}$ and $CR_{\mathrm{ref}}$ are the count rates of the dip block and the reference block, respectively. Thus $\mathrm{depth}=0$ indicates no decrement, while $\mathrm{depth}=1$ corresponds to a full flux drop.

\begin{figure*}
    \centering
    \includegraphics[width=0.49\linewidth]{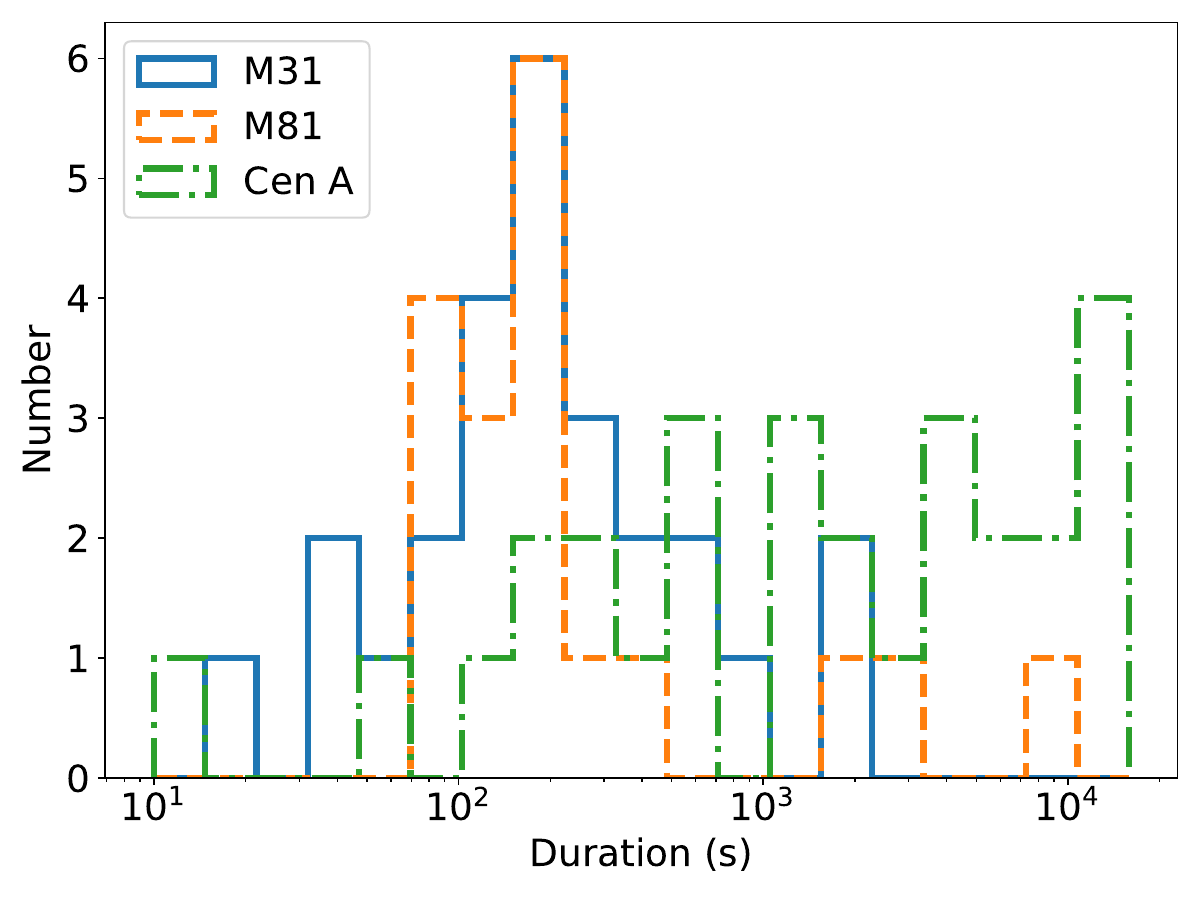}
    \includegraphics[width=0.49\linewidth]{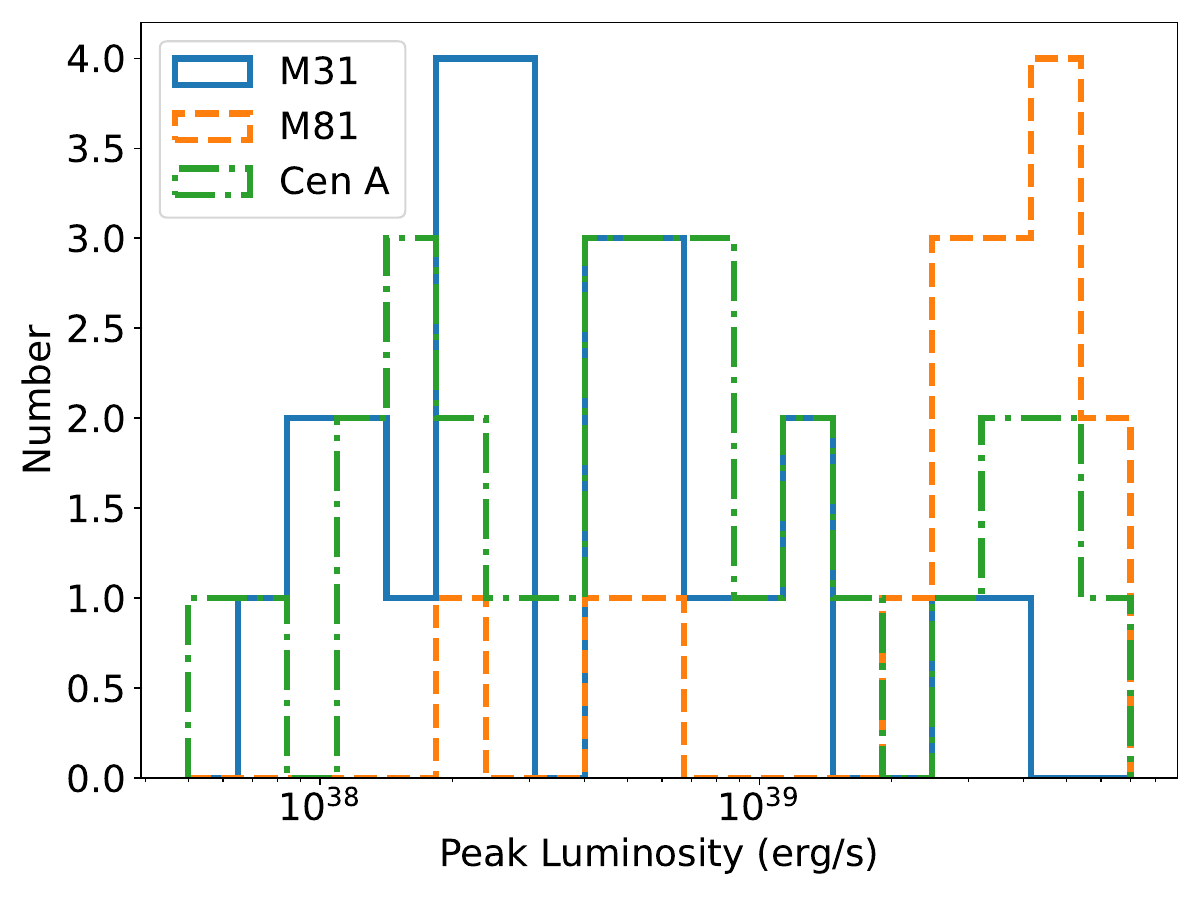}
    \caption{{\it Left}: Histograms of flare durations. {\it Right}: Histograms of flare peak luminosities. Colors denote host galaxies—blue: M31, red: M81, green: Cen A. Durations are defined as the temporal extent of the Bayesian-Blocks–identified flare segment \citep{2013ApJ...764..167S}. For each flare, the corresponding block is subdivided into five equal time bins; the maximum bin count rate is adopted as the peak, and converted to a 0.5–8 keV luminosity(see Section~\ref{sec:data} for spectral assumptions). Distributions are shown with logarithmic binning. 
    }
    \label{fig:hist}
\end{figure*}

The distributions of flare duration and peak luminosity are shown in Figure~\ref{fig:hist}. The flare duration ranges from a few seconds to about two kiloseconds. In M31 and M81 most flares last $10^{2}$–$10^{3}$ s, while in Cen A more than half flares have durations of $10^{3}$–$10^{4}$ s. 
The peak luminosities extend from $10^{37}$--$10^{40}~\rm erg~s^{-1}$. In M31 and Cen A they cluster around $10^{38}$–$10^{39}~\rm erg~s^{-1}$, whereas in M81 they are more frequently $10^{39}$–$10^{40}~\rm erg~s^{-1}$.
The physical nature of the flares are not immediately clear. While a significant fraction of the flares have their peak luminosities close to the Eddington luminosity of an accreting neutron star ($L_{\rm Edd, NS}$), their durations are generally longer than the typical value of Type-I bursts ($\lesssim$100 s, e.g. as cataloged in MINBAR; \citealp{2020ApJS..249...32G}). 
Such flares might be the extragalactic counterpart of an emerging class of unusually long thermonuclear bursts \citep{2023MNRAS.521.3608A}, which have durations of $\sim10^{2-5}$ s and burst energies of $10^{40-42}\rm~erg$. 
Nevertheless, the unusually long thermonuclear bursts still have their peak luminosities consistent with $L_{\rm Edd, NS}$. Therefore, those flares with peak luminosities significantly higher than $L_{\rm Edd, NS}$ are unlikely from normal NS-LMXBs. We speculate that such flares are associated with BH-LMXBs in which a short-term release of the accretion energy temporarily boosts the X-ray emission. Regardless of the detailed physical origin, the {\rm flare duty cycles} of these flares are generally very low, $4.9\times10^{-6}$–$3.6\times10^{-2}$ (mostly below $10^{-2}$; see Table~\ref{tab:flare}), compared with typical values of $\sim 10^{-3}-10^{-1}$ inferred for Galactic type~I X-ray bursters by $DC_{\rm burst}\sim \Delta t_{\rm burst}/t_{\rm rec}$\citep{2020ApJS..249...32G}, where $\Delta t_{\rm burst}$ is the burst duration and $t_{\rm rec}$ is the burst recurrence time. This difference is likely dominated by selection effects: our sources are much more distant, so the effective flux threshold for detecting individual flares is higher and only the brightest events are recovered, biasing the inferred flare duty cycles to lower values.

\begin{figure}
    \centering
    \includegraphics[width=1\linewidth]{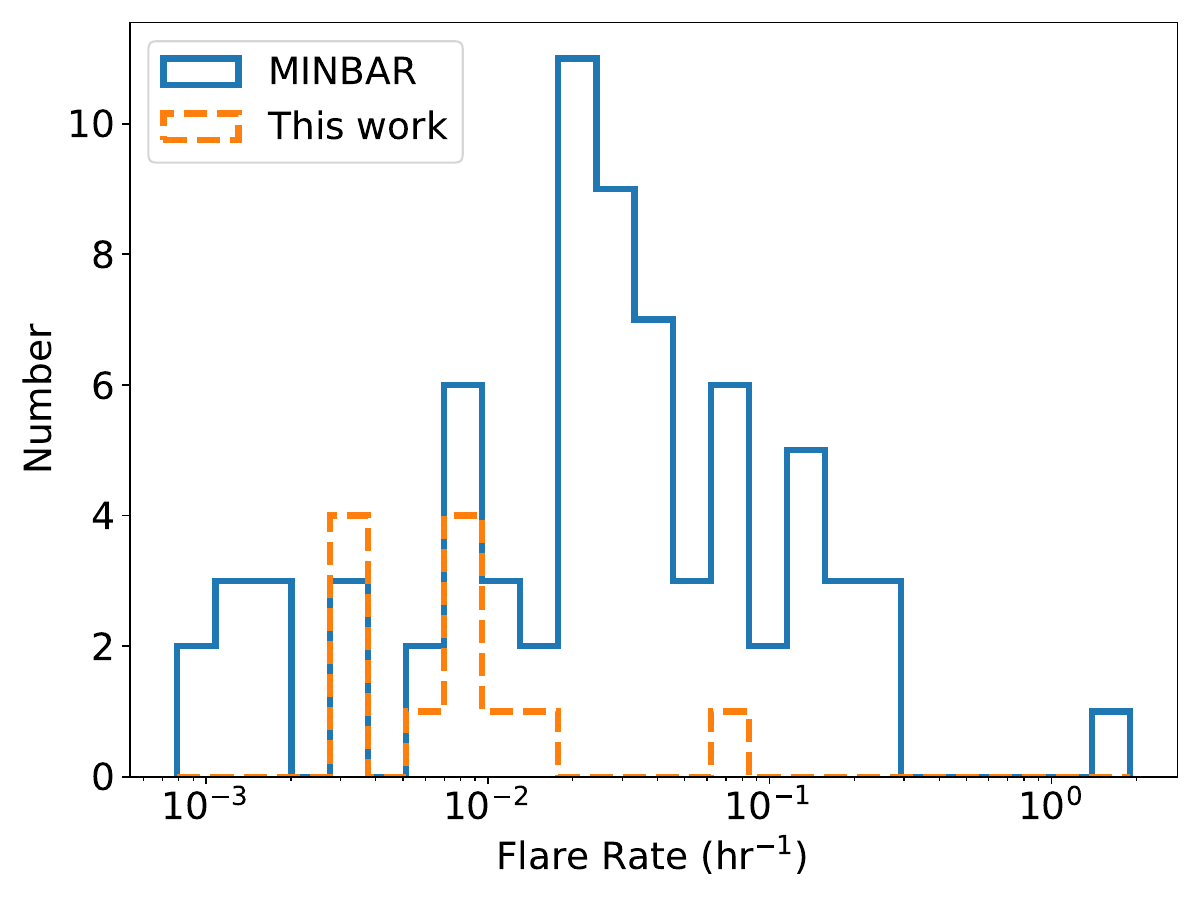}
    \caption{Histogram of the flare rates measured in this work (orange dashed line) and the Type~I burst rates from the MINBAR sample (blue solid line). Only sources with a well-defined rate (i.e., those for which the mean burst or flare rate is constrained rather than a upper limit, and with at least two detected events) are included in both distributions.}

    \label{fig:rate}
\end{figure}

In the representative Galactic Type~I X-ray burst sample from MINBAR, the burst rates span a wide range, from $\sim 4\times10^{-4}$~hr$^{-1}$ to $\sim 1.9$~hr$^{-1}$ \citep{2020ApJS..249...32G}. In contrast, the extragalactic flaring sources identified in this work exhibit flare rates in the interval $\sim 3\times10^{-3}$–$8\times10^{-2}$~hr$^{-1}$. 
To enable a consistent comparison between the two samples, we exclude sources for which only a upper limit on the burst rate is available in MINBAR, as well as sources in our sample with only a single detected flare. The resulting flare–burst rate distributions are shown in Figure~\ref{fig:rate}. 
The difference between the two rate distributions is most likely dominated by selection effects. Because the X-ray binaries in M31, M81, and Cen~A are much more distant than Galactic systems, only relatively bright flares can be detected with {\it Chandra}, and only if they recur with frequencies high enough to appear within our limited observational coverage. As a consequence, systems with very low ($\lesssim 10^{-3}$~hr$^{-1}$) or very high ($\gtrsim 10^{-2}$–$10^{-1}$~hr$^{-1}$) flaring rates are significantly less likely to be identified in our dataset.

The dips have a duration ranging between 0.5--45 ks and a relative depth as low as 15\% and as high as 95\%. For each dip identified by the Bayesian Blocks segmentation, we also compute the dip luminosity directly from the dip block’s net count rate.
We find two phenomenological classes for the tips. The {\it recurrent} dips (e.g. M31~D-Seq~3; Figure~\ref{fig:dips}) appear in more than one observations, although no evidence for periodicity is found within our data coverage. A simple interpretation is geometric obscuration in relatively high–inclination systems, caused by the donor star, structures in the accretion flow such as a stream–disk impact region, a mildly warped/precessing disk, or possibly by intermittent, clumpy outflows near the disk plane \citep{1985SSRv...40..167W,1987A&A...178..137F,2006A&A...445..179D,2016AN....337..512P}.
Recurrent, deep dips with similar phenomenology (flux reductions by factors of 15\% up to near-complete occultation over a substantial fraction like 20\% -- 50\% to the persistent time) are well documented in Galactic dipping LMXBs such as EXO~0748$-$676, MXB 1659$-$298 and 4U~1323$-$62 \citep[e.g.][]{2005A&A...436..195B,2006A&A...445..179D}.
And similar intermittent dipping behaviour has been observed in Galactic LMXB Aql X-1, interpreted as episodes of increased absorption when a thickened region at the outer accretion disc edge sporadically obscures the neutron star, where the dips occupied only 3\% of the total exposure time \citep{2016MNRAS.461.3847G}. The depths of the recurrent dips in this work lie within the reasonable range observed in Galactic samples.

The {\it non-recurrent} dips are events seen only once in single sources in our monitoring observations, which may rise from stochastic line-of-sight obscuration by transient clumps in the disk atmosphere/wind \citep{2005A&A...436..195B,2006A&A...445..179D,2012MNRAS.422L..11P}, or from short-lived changes in the inner accretion flow that temporarily reduce the X-ray output \citep{1997MNRAS.292..679L,2005MNRAS.359..345U}. Their depths and durations fall within the broad range observed in Galactic dipping systems, from relatively shallow, shoulder-like events to deep, short-lived occultations. Given the finite exposure and irregular cadence, we cannot exclude very long orbital or superorbital cycles that fall outside our sensitivity window (orbital period on the order of years; \citealp{2003MNRAS.343.1213C,2009MNRAS.393..139F}).

In order to quantify potential contamination from background AGNs, we also perform a systematic search of flares and dips in the CDFS sources, finding 42 flares and 2 dips.
Then, the expected number of flares due to background AGNs in each of the three galaxy fields can be estimated as

\begin{align}
N_{\mathrm{flare,AGN}} \approx  N_{\mathrm{AGN}}\times \frac{T_{\rm bulge}}{{T_{\mathrm{CDFS}}}} \times \frac{N_{\mathrm{flare,CDFS}}}{N_{\mathrm{CDFS}}},
\end{align}

where 
$N_{\mathrm{AGN}}$ is the statistically expected number of background AGNs, $N_{\mathrm{CDFS}}$ is the total number of X-ray sources in the CDFS,  $N_{\mathrm{flare,CDFS}}$ is the number of flares detected in the CDFS,
and 
$T_{\rm bulge}$ and $T_{\mathrm{CDFS}}$ are the total exposure. 
We restrict our estimate of background contamination for the bulge region, which has a more-or-less uniform exposure. This results in 0.59/27 flares for M31, 0.12/2 flares for M81, and 0.25/21 flares for Cen A, which can be attributed to background AGNs, suggesting that the majority of flares found in the three galaxies are genuine flares associated with their LMXB population. A similar exercise suggests that the number of dips generated from background AGNs is negligible in all three galaxies.

\subsection{Long-term variability}
Following the procedures in Section~\ref{sec:km}, we analyzed  200/456, 41/101, 181/312 bulge sources with at least 5 observations of detections in M31, M81 and Cen A, respectively. 
The mean luminosity E[X] ranges between $3\times10^{35}-8\times10^{38}$ erg~s$^{-1}$. Most sources with E[X] $\lesssim 10^{37}\rm~erg~s^{-1}$ are from M31 due to its proximity (hence a lower sensitivity). 
On the other hand, essentially all sources with E[X] $\gtrsim 2\times10^{38}\rm~erg~s^{-1}$ (i.e. roughly the Eddington limit for an accreting neutron star) are from Cen A.

\begin{figure}
\centering\includegraphics[width=0.99\linewidth]{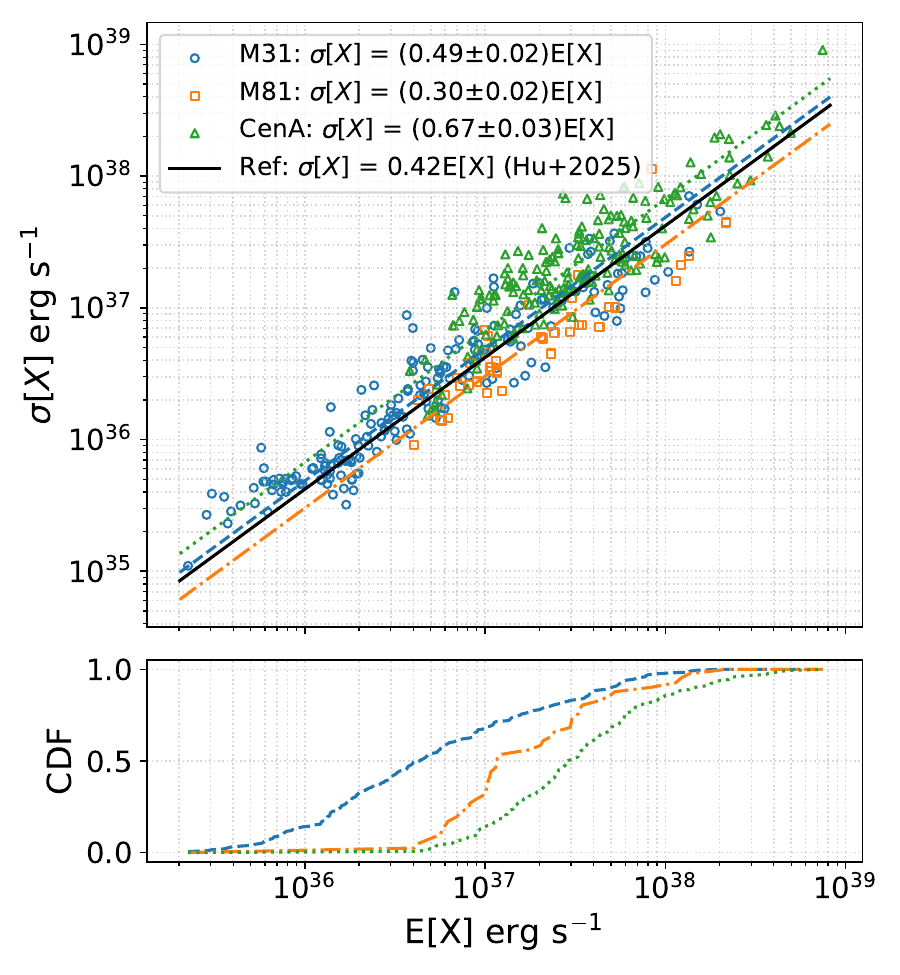}
    \caption{Standard deviation $\sigma[X]$ versus mean luminosity $E[X]$ for central bulge sources in M31 (blue circles), M81 (orange squares) and Cen A (green triangles). Best-fit linear relations ($\sigma[X] \propto E[X]$) are overplotted with the $f$ of $0.49\pm 0.02$, $0.30\pm0.02$, $0.67\pm0.03$ for M31, M81 and Cen A, respectively. The lower panel shows the CDFs of the mean luminosities for the three galaxies. These results demonstrate a linear $\sigma[X]$-$E[X]$ relation consistent with the rms-flux relation observed in galactic XRBs, AGN and high-redshift quasars, as shown by the black line adopted from \citet{2025ApJ...992...82H}.}
    \label{fig:km}
\end{figure}

The standard deviation $\sigma[X]$ of the luminosity distribution and the corresponding mean luminosity are shown in Figure~\ref{fig:km}, with the cumulative distribution function (CDF) displayed in the lower panel (nuclear point sources were excluded to avoid AGN contamination). 
We find that the standard deviation strongly correlates with the mean luminosity. 
Motivated by previous studies of the linear rms–flux relation in Galactic X-ray binaries and AGN \citep[e.g.][]{2001MNRAS.323L..26U,2005MNRAS.359..345U}, we assume the long-term rms–luminosity relation as $\sigma[X] = k \langle E[X] \rangle$. In log–log space, we fit the relation $\log \sigma[X] = \log k + \log \langle E[X] \rangle$ and obtain $k = 0.49\pm0.02$ for M31, $0.30\pm0.02$ for M81, and $0.67\pm0.03$ Cen~A (1$\sigma$ uncertainties).

Within the propagating-fluctuations framework
\citep{1997MNRAS.292..679L,2005MNRAS.359..345U}, the $\sigma$--$E[X]$ linear scaling coefficient measures the typical \emph{fractional rms} of the accretion flow, i.e. $f\simeq\sigma/E[X]$. The numerical differences we obtain therefore quantify genuine changes in variability amplitude across the three bulge populations rather than a distance or calibration effect (the ratio $\sigma/E[X]$ is dimensionless). 
The $f\sim0.30$ in M81 indicates a relatively low fractional rms, consistent with a sample dominated by moderately variable LMXBs. The intermediate $f\sim0.49$ in M31 agrees with the $\sim0.3$--$0.5$ fractional rms commonly reported for Galactic XRBs \citep{2001MNRAS.323L..26U,2005MNRAS.359..345U}, and is consistent with the mixed bulge population of field and GC-LMXBs. The highest $f\sim0.67$ in Cen A suggests that a larger fraction of its bright LMXBs exhibit strong long-term variability, as also observed in early-type hosts \citep{2006MNRAS.371.1903I}. 
Thus, while all three galaxies follow the same linear rms-flux relation, the exact $f$ value traces the dominant accretion states and population characteristics in each environment. 
More recently, \cite{2025ApJ...992...82H} reported a coefficient of $\sim 0.42$ for X-ray-luminous quasars at high redshift. Our measured coefficients are broadly consistent with the range reported in Galactic XRBs, AGN and high-redshift quasar, strengthening the picture that long-term X-ray variability is governed by a common multiplicative process across diverse accreting systems. 

\subsection{Periodic signals}
\label{sec:presult}
For M31, periodicity searches using the GL algorithm have been presented by \citet{2024MNRAS.530.2096Z}, who reported seven significant signals with periods ranging from 463 s to 69411 s. 
Among them, three periodic sources are temporally and positionally coincident with novae and can be understood as spin-orbit modulation of an accreting white dwarf during the thermonuclear outburst. Such sources, however, are generally very soft and of relatively low luminosities, thus they are not expected to be detected in more distant galaxies such as M81 and Cen A. 
The remaining four periodic signals, although with no known optical counterpart, can be attributed to the orbital motion of a putative LMXB, the orbital period of which typically spends the range of a few to a few tens of hours \citep{2023A&C....4200681M}.
The ultra-short period of 463 s, in particular, is revealed by a dip-like feature in the phase-folded light curve. It is suggested that this signal traces an ultra-compact X-ray binary (UCXB) of the shortest known orbital period \citep{2024MNRAS.530.2096Z}, and that the accretor is most likely a stellar-mass black hole\citep{MA2025101181}.

In contrast, no credible (i.e. $P_{\rm GL} > 0.9$) periodic signals were detected in M81 or Cen A when applying the same procedures described in Section~\ref{sec:glmethod}.
At first glance, this is rather surprising, given the fact that several periodic signals are detected in the M31 bulge, which contains a similar number of X-ray sources and has a similar {\it Chandra} exposure.
More quantitatively, there are 337 sources in the bulge region of M31, after excluding foreground stars and CXB sources statistically. This number is 84 and 265 for the bulge of M81 and bulge of Cen A, respectively. Therefore, there should be $\sim1$ and $\sim3$ periodic signals from LMXBs, by a simple scaling of the four signals found for LMXBs in the M31 bulge, neglecting the small difference in the source detection sensitivity among three fields. The latter should only affect the faintest sources, for which no periodic signals would be detected due to the deficiency of photons.

The above estimation suggests that the absence of periodic signals in M81 and Cen A cannot be solely explained by source counts.
Therefore, we further assess the detection rate of the GL algorithm, using artificially generated light curves containing a periodic eclipse, which is the most likely physical cause of a periodic signal. 
Following \cite{2020MNRAS.498.3513B}, we adopt a stepwise light curve to mimic an eclipse, which has a 10\% fraction in phase and a depth of 0.9 relative to an otherwise constant count rate.
Four different periods, 393, 3933, 7933 and 13933 s, and eight different count rates 0.002, 0.004, 0.006, 0.008, 0.012, 0.018, 0.020, 0.024 $\rm cts\ s^{-1}$, are assumed. For each combination of period and count rate, a total of 1000 light curves are simulated taking into account the Poisson fluctuation. For each simulation set, we use a multi-epoch, gap-separated sampling cadence that mirrors the ACIS-S observations of M81.

\begin{figure}
    \centering
    \includegraphics[width=0.99\linewidth]{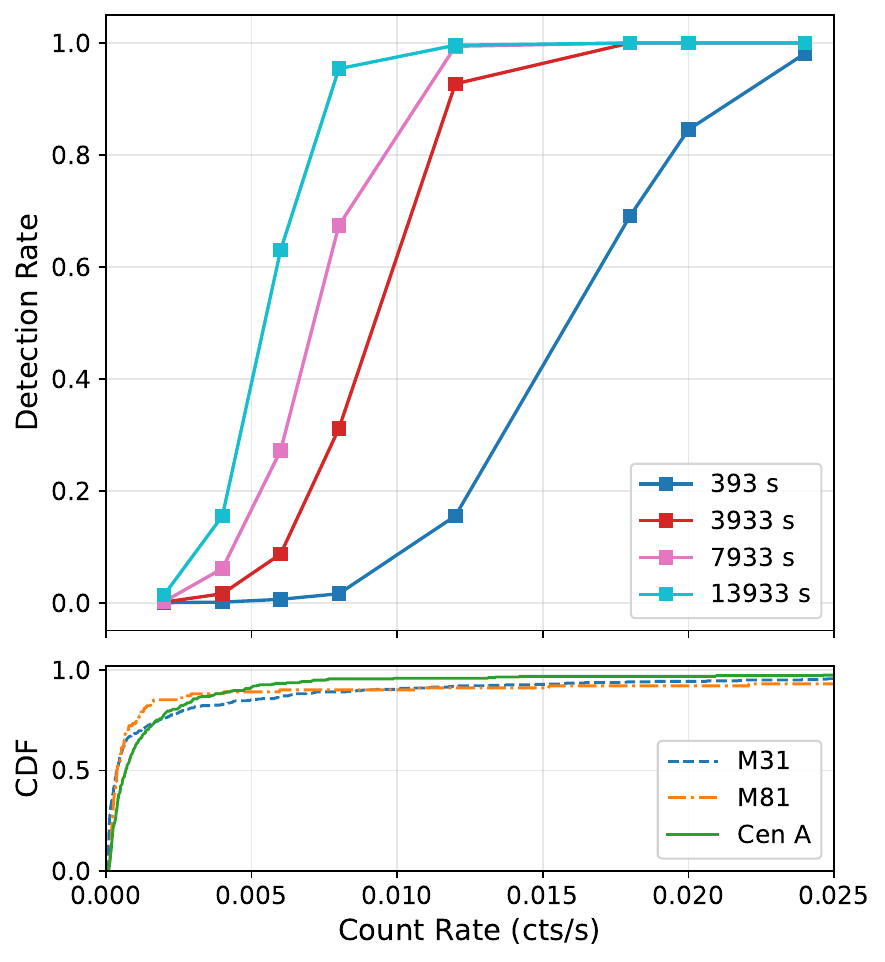}
    \caption{Detection rate of periodic signals versus source count rate and trial period by the GL algorithm. As expected, the efficiency increases with count rate. At a fixed count rate, the efficiency rises with period. The lower panel shows the CDFs of the mean source count rates for the three galaxies. }
    \label{fig:detectionrate}
\end{figure}

These light curves are then fed to the GL algorithm in the same manner as for the real light curves. 
The resulting detection rates, defined as the fraction of the simulated light curve for which the period of the eclipsing signal is recovered (i.e. a reported period with $P_{\rm GL} > 0.9$ and within few per cents of the input period), are illustrated in Figure~\ref{fig:detectionrate}. As expected, the detection rate increases with source count rate, while at a fixed count rate, the detection efficiency increases with increasing period. 
From the predicted detection efficiency and the observed distribution of source count rates in the three galaxies, our period search is essentially complete for sources with count rates $\gtrsim 0.018~\mathrm{cts~s^{-1}}$ and trial periods between $\sim 1-20~\mathrm{ks}$. In contrast, a substantial fraction of fainter sources would have their periodic signals buried in the noise. We also note that the irregular dipping seen in several sources  (Section~\ref{sec:flare}) could have arisen from periodicities longer than our $20~\mathrm{ks}$ search window (e.g., orbital or superorbital modulations), which our analysis would not recover.

Guided by population-synthesis models of LMXBs including NS systems \citep{2019MNRAS.483.5595V} and BH systems \citep{2020ApJ...898..143S}, higher-luminosity LMXBs tend to have longer orbital periods, whereas shorter-period systems preferentially occupy the lower-luminosity regime, with UCXBs being a notable exception. In the bulge regions of M81 and Cen~A, the $\sim$100 count practical limit for periodic signal detection by the GL algorithm corresponds to a luminosity threshold of $\sim 10^{37}\ \mathrm{erg\,s^{-1}}$, while for M31 the corresponding threshold is $\sim 10^{36}\ \mathrm{erg\,s^{-1}}$ owing to its closer distance. The higher thresholds for M81 and Cen~A push many short-period (and thus typically lower-luminosity) systems below our period-search completeness limit. UCXBs, by contrast, can be relatively luminous at short periods. We did detect one UCXB candidate in M31 \citep{2024MNRAS.530.2096Z,MA2025101181}, but none in M81 or Cen~A. Consequently, in M81 and Cen~A the LMXBs having short orbital periods are the ones most likely to fall below our detection sensitivity. This selection effect, combined with the rarity of luminous short-period UCXBs, naturally lowers the expected yield of coherent detections in M81 and Cen~A relative to M31.
On the other hand, although Cen~A and M81 harbor more luminous LMXBs ($L_{\rm X} \gtrsim 10^{38}\rm~erg~s^{-1}$) than M31 owing to their higher enclosed stellar masses, such systems generally have orbital periods greater than $\sim10$ hrs \citep{2019MNRAS.483.5595V,2020ApJ...898..143S}, which cannot be efficiently detected with the current X-ray data.

For a rough quantitative estimate of the total number of detectable periodic signals, we adopt that $\sim 50\%$ of LMXBs have $P_{\rm orb}<20~\mathrm{ks}$ \citep{2019MNRAS.483.5595V,2020ApJ...898..143S} and fold this fraction through the above luminosity thresholds and source counts in each galaxy. 
This implies about 168, 42, and 132 short-period ($<20~\mathrm{ks}$) bulge LMXBs in M31, M81, and Cen~A, respectively, which in principle could be identified with the Gregory-Loredo algorithm. A further geometric factor reduces the observable fraction: only systems with inclination $60^\circ \lesssim i \lesssim 80^\circ$ are expected to show dips or eclipses \citep{1987A&A...178..137F}, corresponding to an order-of-magnitude fraction of $\sim20\%$ for randomly oriented binaries. Combining the inclination selection with our empirically calibrated detection-efficiency–versus–count-rate curve yields final expectations of $\approx4$ (M31), $\approx1$ (M81), and $\approx2$ (Cen~A) detections for $P_{\rm orb}<20~\mathrm{ks}$. As a cross-check, M31 has four previously reported LMXB periodicities below $20~\mathrm{ks}$ \citep{2024MNRAS.530.2096Z}, in line with our expectations at its depth. The observed yields are consistent with these expectations when stochastic fluctuations, low flare duty cycles, and small-number statistics are considered.

For the CDFS, \cite{2022MNRAS.509.3504B} has searched for periodicity using the GL algorithm, which, resulted no significant detection when combining the 7 Ms exposure. 
This suggests that periodic X-ray signals are rather rare in background AGNs under a typical {\it Chandra} exposure, an empirical result now gaining support from the non-detections in the M81 and Cen A fields.

\section{Summary}\label{sec:sum}
Based on 20-year \textit{Chandra} archival data, we have conducted a systematic timing analysis of several hundred X-ray sources in M31, M81, and Cen~A (NGC 5128), mostly LMXBs, focusing on their periodic and aperiodic variability. Our main results are:

\begin{enumerate}
\item \textbf{Flare census and properties.} We identified flares in 24 sources in M31, 5 in M81, and 26 in Cen~A, with several sources exhibiting recurrent flaring. Flare durations range from $\sim$10~s to a few $10^{4}$~s, and peak luminosities span $10^{37}$--$10^{40}\ \mathrm{erg\,s^{-1}}$. Flare duty cycles are generally low, $4.9\times10^{-6}$--$3.5\times10^{-2}$, indicating brief episodes superposed on long quiescent intervals. 

\item \textbf{Foreground stellar flares.} We additionally identified flares in 13 foreground stars, which is an unexpected but informative byproduct of our extragalactic survey. Five of them show recurrent events, providing a uniform-\textit{Chandra} baseline on stellar flare energetics.

\item \textbf{Dip events.} We detected dips in 8 sources in M31, 1 in M81, and 5 in Cen~A; two sources show repeating dips. The mixture of recurrent and isolated dips is naturally explained by time-variable geometric obscuration in high-inclination systems and by transient clumps or short-lived structural changes in the accretion flow.

\item \textbf{Long-term variability (rms--flux relation).} Across multi-epoch baselines, the standard deviation $\sigma[X]$ correlates linearly with the mean luminosity $E[X]$, with best-fit linear scaling coefficient of 0.49 (M31), 0.30 (M81), and 0.67 (Cen~A). These differing linear scaling coefficient point to galaxy-to-galaxy diversity in variability amplitude at fixed mean luminosity. 

\item \textbf{Periodic signals.} Within our sensitivity limits, no statistically significant coherent periodicity is detected in M81 or Cen~A. 

\end{enumerate}
Taken together, the prevalence of short-duty-cycle flares, the coexistence of recurrent and isolated dips, and the galaxy-dependent rms--flux linear scaling coefficient support a picture in which (i) Stochastic accretion-rate fluctuations within the inner accretion flow modulate the X-ray luminosity on $\sim$10–$10^{4}$ s, consistent with dynamical–thermal–viscous timescales; (ii) at high inclinations, time-variable geometric obscuration (e.g. stream–disk impact bulges, mild warps, or clumpy equatorial outflows) naturally produces dips without requiring a strictly phase-locked clock. Conducting this study in external galaxies provides a distance-calibrated, population-level view that is less affected by line-of-sight and selection biases inherent to the Milky Way, establishing extragalactic view of flare duty cycles, dip phenomenology, and rms-flux scalings for bulge-dominated environments. These empirical results promise to strengthen constraints on accretion physics across different host potentials and stellar populations. 
It would be beneficial to extend the analysis to more external galaxies utilizing the rich {\it Chandra} inventory.

\section*{Acknowledgements}

This work is supported by the National Natural Science Foundation of China (grant 12225302), the Fundamental Research Funds for the Central Universities (grant KG202502) and the CNSA program D050102. 
The authors wish to thank Prof. Yong Shao and Prof. Natasha Ivanova for helpful discussions on the orbital period distribution of LMXBs.
\section*{Data Availability}
The data underlying this paper will be shared on reasonable request to the corresponding author. The Chandra data used in this paper are available in the Chandra Data Archive (https://cxc.harvard.edu/cda/) by searching the Observation ID listed in Tables~\ref{tab:obsCenA} and \ref{tab:obsM81} in the Search and Retrieval interface, ChaSeR (https://cda.harvard.edu/chaser/).



\bibliographystyle{mnras}
\bibliography{example} 




\appendix
\onecolumn

\section{{\it Chandra} observations of M81 and Cen A}
Basic information of the {\it Chandra} observations of Cen A and M81 utilized in this work are given in Tables \ref{tab:obsCenA} and \ref{tab:obsM81}.
\begin{longtable}{cccccc}
\caption{{\it Chandra} Observations of Cen A used in this work}\label{tab:obsCenA}\\
\hline
Obs ID	&	Instrument	&	RA	&	Dec	&	Start Time	&	Exposure (ks)	\\
\hline\hline
7797	&	ACIS-I	&	13 25 19.15	&	-43 02 42.40	&	2007/3/22 08:59	&	98.18 	\\
7798	&	ACIS-I	&	13 25 51.80	&	-43 00 04.43	&	2007/3/27 09:53	&	92.05 	\\
7799	&	ACIS-I	&	13 25 51.80	&	-43 00 04.43	&	2007/3/30 02:32	&	96.05 	\\
7800	&	ACIS-I	&	13 25 46.00	&	-42 58 14.58	&	2007/4/17 15:00	&	92.04 	\\
8489	&	ACIS-I	&	13 25 32.79	&	-43 01 35.13	&	2007/5/8 18:41	&	95.18 	\\
8490	&	ACIS-I	&	13 25 18.79	&	-43 03 01.72	&	2007/5/30 02:01	&	95.68 	\\
10723	&	ACIS-I	&	13 25 49.66	&	-42 59 14.74	&	2009/1/4 12:32	&	5.15 	\\
10724	&	ACIS-I	&	13 25 28.14	&	-43 04 31.70	&	2009/3/5 07:51	&	5.17 	\\
10725	&	ACIS-I	&	13 25 34.95	&	-43 03 26.79	&	2009/4/26 21:55	&	5.04 	\\
10726	&	ACIS-I	&	13 25 51.08	&	-42 57 36.32	&	2009/6/21 07:03	&	5.14 	\\
11846	&	ACIS-I	&	13 25 39.40	&	-43 03 47.40	&	2010/4/26 21:14	&	4.75 	\\
11847	&	ACIS-I	&	13 25 12.80	&	-42 59 51.20	&	2010/9/16 17:35	&	5.05 	\\
12155	&	ACIS-I	&	13 25 47.69	&	-42 58 45.13	&	2010/12/22 12:15	&	5.05 	\\
12156	&	ACIS-I	&	13 25 17.55	&	-43 01 54.14	&	2011/6/22 20:11	&	5.06 	\\
13303	&	ACIS-I	&	13 25 44.29	&	-43 01 59.89	&	2012/4/14 04:38	&	5.58 	\\
13304	&	ACIS-I	&	13 25 17.73	&	-43 01 02.87	&	2012/8/29 16:20	&	5.08 	\\
15294	&	ACIS-I	&	13 25 27.00	&	-43 01 08.00	&	2013/4/5 23:30	&	5.08 	\\
15295	&	ACIS-I	&	13 25 27.00	&	-43 01 08.00	&	2013/8/31 21:57	&	5.58 	\\
16276	&	ACIS-I	&	13 25 39.17	&	-43 03 20.00	&	2014/4/24 15:42	&	5.09 	\\
16277	&	ACIS-I	&	13 25 07.69	&	-43 01 14.04	&	2014/9/8 04:47	&	5.08 	\\
17471	&	ACIS-I	&	13 25 24.99	&	-43 02 24.85	&	2015/3/14 02:05	&	5.08 	\\
17472	&	ACIS-I	&	13 25 27.60	&	-43 01 09.00	&	2015/9/21 02:53	&	4.40 	\\
18461	&	ACIS-I	&	13 25 31.52	&	-43 02 23.50	&	2016/4/23 12:33	&	5.19 	\\
18462	&	ACIS-I	&	13 25 24.69	&	-42 59 55.34	&	2016/9/20 07:25	&	5.07 	\\
19747	&	ACIS-I	&	13 25 27.60	&	-43 01 09.00	&	2017/5/15 21:39	&	5.07 	\\
19748	&	ACIS-I	&	13 25 43.60	&	-42 58 16.50	&	2017/12/12 04:21	&	5.06 	\\
20752	&	ACIS-I	&	13 25 25.66	&	-43 01 58.39	&	2018/3/8 07:30	&	5.06 	\\
20753	&	ACIS-I	&	13 25 27.60	&	-43 01 09.00	&	2018/8/31 22:20	&	5.06 	\\
21698	&	ACIS-I	&	13 25 32.17	&	-43 03 35.40	&	2019/4/29 07:33	&	5.07 	\\
21699	&	ACIS-I	&	13 25 27.62	&	-43 01 08.80	&	2019/9/10 00:06	&	5.07 	\\
22714	&	ACIS-I	&	13 25 34.40	&	-43 03 41.80	&	2020/5/3 12:38	&	4.95 	\\
22715	&	ACIS-I	&	13 25 21.17	&	-43 00 47.30	&	2020/9/14 02:29	&	4.96 	\\
26038	&	ACIS-I	&	13 25 57.00	&	-43 00 43.00	&	2022/3/6 16:05	&	4.96 	\\
26039	&	ACIS-I	&	13 25 16.00	&	-42 58 01.00	&	2022/9/20 00:18	&	5.17 	\\
2978	&	ACIS-S	&	13 25 28.69	&	-43 00 59.70	&	2002/9/3 2:42	&	45.17 	\\
3965	&	ACIS-S	&	13 25 28.70	&	-43 00 59.70	&	2003/9/14 13:44	&	50.17 	\\
10722	&	ACIS-S	&	13 25 27.60	&	-43 01 09.00	&	2009/9/8 20:05	&	50.03 	\\
17890	&	ACIS-S	&	13 25 24.11	&	-43 00 57.50	&	2016/2/24 22:45	&	10.06 	\\
17891	&	ACIS-S	&	13 25 24.11	&	-43 00 57.50	&	2016/3/22 2:17	&	10.05 	\\
19521	&	ACIS-S	&	13 25 27.60	&	-43 01 09.00	&	2017/9/17 23:25	&	15.06 	\\
20794	&	ACIS-S	&	13 25 27.60	&	-43 01 09.00	&	2017/9/19 06:57	&	108.65 	\\
20585	&	ACIS-S	&	13 25 52.70	&	-43 05 46.00	&	2019/4/22 19:47	&	50.74 	\\
22189	&	ACIS-S	&	13 25 52.70	&	-43 05 46.00	&	2019/4/24 12:52	&	30.06 	\\
\hline
\end{longtable}
\begin{longtable}{cccccc}
\caption{{\it Chandra} Observations of M81 used in this work}\label{tab:obsM81}\\
\hline
ObsID	&	Instrument	&	RA	&	Dec	&	Start Time		&	Exposure (ks)	\\
    \hline\hline
18052	&	ACIS-I	&	09 56 03.80	&	+68 53 08.5	&	2016-02-01 09:51:24		&	20.47 	\\
18054	&	ACIS-I	&	09 54 12.70	&	+69 01 36.8	&	2016-06-21 08:35:49		&	66.58 	\\
18875	&	ACIS-I	&	09 54 12.70	&	+69 01 36.8	&	2016-06-24 15:20:01		&	32.79 	\\
18047	&	ACIS-I	&	09 56 44.90	&	+69 06 33.0	&	2016-07-04 14:36:06		&	101.76 	\\
18817	&	ACIS-I	&	09 56 03.80	&	+68 53 08.5	&	2016-08-07 23:28:52		&	40.07 	\\
19685	&	ACIS-I	&	09 56 03.80	&	+68 53 08.5	&	2016-08-08 16:32:46		&	37.07 	\\
18049	&	ACIS-I	&	09 54 25.40	&	+69 14 46.4	&	2016-08-15 01:37:50		&	20.21 	\\
19688	&	ACIS-I	&	09 54 25.40	&	+69 14 46.4	&	2016-08-16 12:42:30		&	63.87 	\\
18048	&	ACIS-I	&	09 56 44.90	&	+69 06 33.0	&	2017-01-08 10:29:32		&	70.07 	\\
19981	&	ACIS-I	&	09 56 44.90	&	+69 06 33.0	&	2017-01-11 12:17:36		&	30.05 	\\
19982	&	ACIS-I	&	09 54 12.70	&	+69 01 36.8	&	2017-01-12 06:03:53		&	68.39 	\\
18053	&	ACIS-I	&	09 54 12.70	&	+69 01 36.8	&	2017-01-15 06:51:23		&	29.98 	\\
18051	&	ACIS-I	&	09 56 03.80	&	+68 53 08.5	&	2017-01-24 06:52:36		&	37.06 	\\
18050	&	ACIS-I	&	09 54 25.40	&	+69 14 46.4	&	2017-01-24 17:42:50		&	54.77 	\\
19993	&	ACIS-I	&	09 56 03.80	&	+68 53 08.5	&	2017-01-26 21:51:04		&	25.08 	\\
19992	&	ACIS-I	&	09 56 03.80	&	+68 53 08.5	&	2017-01-28 04:05:58		&	35.06 	\\
19991	&	ACIS-I	&	09 54 25.40	&	+69 14 46.4	&	2017-01-29 13:45:44		&	54.06 	\\
21384	&	ACIS-I	&	09 55 24.20	&	+69 09 57.0	&	2019-09-04 21:13:33		&	42.53 	\\
390	&	ACIS-S	&	09 55 33.20	&	+69 03 55.0	&	2000-03-21 00:41:43		&	2.40 	\\
735	&	ACIS-S	&	09 55 25.00	&	+69 01 12.0	&	2000-05-07 05:31:17		&	50.65 	\\
4751	&	ACIS-S	&	09 57 54.30	&	+69 03 46.4	&	2004-02-07 10:54:05		&	5.08 	\\
4752	&	ACIS-S	&	09 57 54.30	&	+69 03 46.4	&	2004-03-25 10:02:51		&	5.15 	\\
5935	&	ACIS-S	&	09 55 33.20	&	+69 03 55.1	&	2005-05-26 03:46:22		&	11.12 	\\
5936	&	ACIS-S	&	09 55 33.20	&	+69 03 55.1	&	2005-05-28 19:54:20		&	11.55 	\\
5937	&	ACIS-S	&	09 55 33.20	&	+69 03 55.1	&	2005-06-01 08:31:48		&	12.16 	\\
5938	&	ACIS-S	&	09 55 33.20	&	+69 03 55.1	&	2005-06-03 22:48:06		&	11.96 	\\
5939	&	ACIS-S	&	09 55 33.20	&	+69 03 55.1	&	2005-06-06 15:16:37		&	11.95 	\\
5940	&	ACIS-S	&	09 55 33.20	&	+69 03 55.1	&	2005-06-09 06:53:16		&	12.12 	\\
5941	&	ACIS-S	&	09 55 33.20	&	+69 03 55.1	&	2005-06-11 21:22:11		&	11.96 	\\
5942	&	ACIS-S	&	09 55 33.20	&	+69 03 55.1	&	2005-06-15 01:27:29		&	12.10 	\\
5943	&	ACIS-S	&	09 55 33.20	&	+69 03 55.1	&	2005-06-18 11:33:20		&	12.16 	\\
5944	&	ACIS-S	&	09 55 33.20	&	+69 03 55.1	&	2005-06-21 05:19:56		&	11.96 	\\
5945	&	ACIS-S	&	09 55 33.20	&	+69 03 55.1	&	2005-06-24 06:04:49		&	11.72 	\\
5946	&	ACIS-S	&	09 55 33.20	&	+69 03 55.1	&	2005-06-26 21:42:11		&	12.17 	\\
5947	&	ACIS-S	&	09 55 33.20	&	+69 03 55.1	&	2005-06-29 13:33:37		&	10.83 	\\
5948	&	ACIS-S	&	09 55 33.20	&	+69 03 55.1	&	2005-07-03 01:33:34		&	12.18 	\\
5949	&	ACIS-S	&	09 55 33.20	&	+69 03 55.1	&	2005-07-06 07:49:39		&	12.17 	\\
9805	&	ACIS-S	&	09 55 26.00	&	+69 04 34.8	&	2007-12-21 18:41:10		&	5.17 	\\
9122	&	ACIS-S	&	09 55 24.80	&	+69 01 13.7	&	2008-02-01 00:23:54		&	10.03 	\\
9540	&	ACIS-S	&	09 57 32.00	&	+69 02 45.0	&	2008-08-24 09:50:36		&	26.05 	\\
12301	&	ACIS-S	&	09 55 24.80	&	+69 01 13.7	&	2010-08-18 13:52:07		&	79.05 	\\
13728	&	ACIS-S	&	09 57 53.30	&	+69 03 48.1	&	2012-07-20 03:47:53		&	15.07 	\\
14471	&	ACIS-S	&	09 57 53.30	&	+69 03 48.1	&	2012-09-24 05:23:30		&	14.07 	\\
26421	&	ACIS-S	&	09 55 20.70	&	+69 04 00.9	&	2022-06-04 11:27:37		&	10.08 	\\
\hline
\end{longtable}

\section{Result of flare and dip searching}
{\fontsize{7pt}{12pt}\selectfont
\setlength{\tabcolsep}{1.2pt}
\renewcommand{\arraystretch}{0.85}
\begin{ThreePartTable}
\begin{TableNotes}[flushleft]
\footnotesize
\item \textit{Notes.—}
(1) Galaxy: host galaxy. 
(2) Seq: flare sequence ID; labels correspond to those used in Figures~\ref{fig:acis}. 
(3) CXO name: designation in the \textit{Chandra} Source Catalog (v2.1.1).
(4)–(5) RA/Dec: J2000 coordinates in decimal degrees of the X-ray source centroid. 
(6) ObsID: \textit{Chandra} observation identifier. 
(7) Start Time (MJD): start time of the flare block identified by the Bayesian-Blocks algorithm \citep{2013ApJ...764..167S}. 
(8) Duration (s): temporal extent of the flare block. 
(9) ${CR_{\rm peak}}$ (cts\,s$^{-1}$): peak count rate within the flare, obtained by subdividing the flare block into five equal time bins and taking the maximum bin value (Section~\ref{sec:data}). 
(10) ${CR_{\rm ref}}$ (cts\,s$^{-1}$): reference (non-flare) count rate measured from the reference quiescent block. 
(11) $\gamma$: ratio of the persistent to burst flux, defined here as $CR_{\rm ref}/CR_{\rm peak}$, where $CR_{\rm ref}$ and $CR_{\rm peak}$ are 
the persistent and peak count rates, respectively.
(12) ${L_{\rm peak}}$ ($\rm 10^{38}~erg~ s^{-1}$): absorbed 0.5–8\,keV luminosity converted from $CR_{\rm peak}$ using the instrumental response of the parent observation, the spectral assumption described in Section~\ref{sec:data}.
(13) ${E_{\rm flare}}$ ($\rm 10^{40}~erg$): flare energy (fluence) estimated by summing the flare excess above the reference level, $E_{\rm flare}= (L-L_{\rm ref})\Delta t_{\rm flare} \approx (CR-CR_{\rm ref})\eta\Delta t_{\rm flare}$, where $\eta$ is the count-to-luminosity conversion factor for the observation. 
(14) $r_{\rm flare}$: flare rate, estimated as $r_{\rm flare} = N_{\rm flare} / t_{\rm total}$, where $N_{\rm flare}$ is the number of detected flares and $t_{\rm total}$ is the total on-source exposure time.
(15) Flare Duty Cycle: fraction of recurrence time occupied by flare blocks for the source, i.e. $\mathrm{DC_{flare}}=\Delta t_{{\rm flare}}/t_{\rm rec}$, where $t_{\rm rec}=1/r_{\rm flare}$.
All the quoted uncertainties are $1\sigma$. 
\item A machine-readable version of this table is available in the online journal.
\end{TableNotes}
\begin{longtable}{ccccccccccccccc}
\caption{Flares in M31, M81 and Cen A}\label{tab:flare}\\
\toprule
Galaxy	&	Seq	&	CXO name	&	Ra	&	Dec	&	ObsID	&	Start Time	&	Duration	&	$CR_{\rm peak}$	&	$CR_{\rm ref}$	&	$\gamma$&	$L_{\rm peak}$	&	$E_{\rm flare}$	&	$r_{\rm flare}$	&	Flare Duty Cycle	\\	
	&		&		&	deg	&	deg	&		&	MJD	&	s	&	cts s$^{-1}$	&	cts s$^{-1}$	&	&	$\rm 10^{38}~erg~ s^{-1}$	&	$\rm 10^{40}~erg$	&	$\rm hr^{-1}$	&		\\	
	(1)	&(2)&(3)&(4)&(5)&(6)&(7)&(8)&(9)&(10)&(11)&(12)&(13)&(14)&(15)\\
    \midrule
    \endfirsthead

\midrule
	\multicolumn{15}{l}{\small \textit{(continued)}}\\
	\midrule
Galaxy	&	Seq	&	CXO name	&	Ra	&	Dec	&	ObsID	&	Start Time	&	Duration	&	$CR_{\rm peak}$	&	$CR_{\rm ref}$	&	$\gamma$&	$L_{\rm peak}$	&	$E_{\rm flare}$	&	$r_{\rm flare}$	&	Flare Duty Cycle	\\	
	&		&		&	deg	&	deg	&		&	MJD	&	s	&	cts s$^{-1}$	&	cts s$^{-1}$	&	&	$\rm 10^{38}~erg~ s^{-1}$	&	$\rm 10^{40}~erg$	&	$\rm hr^{-1}$	&		\\	
	(1)	&(2)&(3)&(4)&(5)&(6)&(7)&(8)&(9)&(10)&(11)&(12)&(13)&(14)&(15)\\
	\midrule
	\endhead
	
	\midrule
	\multicolumn{15}{r}{\small \textit{(to be continued)}}\\
	\endfoot
	
	\bottomrule
	\insertTableNotes 
	\endlastfoot

M31	&	1	&	2CXO J004303.1+411015	&	10.76293 	&	41.17101 	&	15267	&	56156.1341 	&	142.27 	&	0.14	$\pm$	0.07	&	0.0048	$\pm$	0.0009	&	3.41E-02	&	3	$\pm$	1	&	1.8	$\pm$	0.7	&	4.00E-03	&	1.58E-04	\\
M31	&	2	&	2CXO J004247.8+411113	&	10.69915 	&	41.18718 	&	14196	&	56228.8023 	&	178.74 	&	0.14	$\pm$	0.06	&	0.0275	$\pm$	0.0008	&	1.97E-01	&	3	$\pm$	1	&	2.6	$\pm$	0.9	&	7.00E-03	&	3.48E-04	\\
M31	&	3	&	2CXO J004228.2+411222	&	10.61784 	&	41.20644 	&	4722	&	53309.1322 	&	20.03 	&	0.7	$\pm$	0.4	&	0.036	$\pm$	0.003	&	4.83E-02	&	13	$\pm$	7	&	0.9	$\pm$	0.4	&	8.54E-03	&	4.75E-05	\\
M31	&	3	&	2CXO J004228.2+411222	&	10.61784 	&	41.20644 	&	13825	&	56079.8391 	&	334.24 	&	0.16	$\pm$	0.05	&	0.052	$\pm$	0.001	&	3.14E-01	&	2	$\pm$	0.6	&	2.6	$\pm$	0.8	&	8.54E-03	&	7.93E-04	\\
M31	&	4	&	2CXO J004215.1+411234	&	10.56306 	&	41.20964 	&	1575	&	52187.4356 	&	278.75 	&	0.13	$\pm$	0.05	&	0.032	$\pm$	0.001	&	2.55E-01	&	1.7	$\pm$	0.6	&	2.2	$\pm$	0.7	&	1.14E-02	&	8.85E-04	\\
M31	&	4	&	2CXO J004215.1+411234	&	10.56306 	&	41.20964 	&	16295	&	56832.2579 	&	482.18 	&	0.07	$\pm$	0.03	&	0.01	$\pm$	0.002	&	1.43E-01	&	1.9	$\pm$	0.7	&	4	$\pm$	1	&	1.14E-02	&	1.53E-03	\\
M31	&	5	&	2CXO J004308.6+411248	&	10.78598 	&	41.21345 	&	13837	&	55976.6863 	&	182.27 	&	0.05	$\pm$	0.04	&	0.0024	$\pm$	0.0009	&	4.37E-02	&	1.1	$\pm$	0.8	&	1.3	$\pm$	0.6	&	4.28E-03	&	2.16E-04	\\
M31	&	6	&	2CXO J004232.0+411314	&	10.63362 	&	41.22069 	&	309	&	51696.0893 	&	92.38 	&	0.4	$\pm$	0.1	&	0.055	$\pm$	0.004	&	1.45E-01	&	2.2	$\pm$	0.8	&	0.8	$\pm$	0.3	&	3.64E-03	&	9.35E-05	\\
M31	&	7	&	2CXO J004222.4+411334	&	10.59341 	&	41.22616 	&	1854	&	51922.5021 	&	158.83 	&	0.16	$\pm$	0.07	&	0.033	$\pm$	0.003	&	2.06E-01	&	1.4	$\pm$	0.6	&	1	$\pm$	0.4	&	4.58E-03	&	2.02E-04	\\
M31	&	8	&	2CXO J004210.3+411509	&	10.54289 	&	41.25279 	&	8191	&	54269.2997 	&	36.54 	&	0.7	$\pm$	0.3	&	0.005	$\pm$	0.001	&	7.16E-03	&	13	$\pm$	6	&	2.2	$\pm$	0.6	&	5.30E-03	&	5.38E-05	\\
M31	&	9	&	2CXO J004248.5+411521	&	10.70219 	&	41.25589 	&	7068	&	54254.0381 	&	203.42 	&	0.17	$\pm$	0.07	&	0.019	$\pm$	0.001	&	1.10E-01	&	6	$\pm$	2	&	6	$\pm$	2	&	3.61E-03	&	2.04E-04	\\
M31	&	10	&	2CXO J004302.9+411522	&	10.76222 	&	41.25628 	&	1575	&	52187.4850 	&	4.89 	&	5	$\pm$	2	&	0.019	$\pm$	0.0007	&	3.72E-03	&	30	$\pm$	10	&	0.3	$\pm$	0.1	&	3.61E-03	&	4.91E-06	\\
M31	&	11	&	2CXO J004258.3+411529	&	10.74298 	&	41.25812 	&	10552	&	54869.5003 	&	228.26 	&	0.07	$\pm$	0.04	&	0.0013	$\pm$	0.0006	&	1.98E-02	&	1.1	$\pm$	0.6	&	0.9	$\pm$	0.4	&	3.61E-03	&	2.29E-04	\\
M31	&	12	&	2CXO J004252.5+411539	&	10.71884 	&	41.26110 	&	305	&	51523.2897 	&	194.47 	&	0.1	$\pm$	0.05	&	0.005	$\pm$	0.001	&	5.15E-02	&	3	$\pm$	1	&	2.4	$\pm$	0.9	&	3.61E-03	&	1.95E-04	\\
M31	&	13	&	2CXO J004246.1+411543	&	10.69230 	&	41.26200 	&	13826	&	56084.8370 	&	547.03 	&	0.1	$\pm$	0.03	&	0.0024	$\pm$	0.0003	&	2.39E-02	&	1.1	$\pm$	0.3	&	4.2	$\pm$	0.7	&	3.72E-03	&	5.66E-04	\\
M31	&	14	&	2CXO J004239.9+411547	&	10.66663 	&	41.26322 	&	7064	&	54074.1745 	&	121.74 	&	0.12	$\pm$	0.07	&	0.0026	$\pm$	0.0003	&	2.11E-02	&	4	$\pm$	3	&	2	$\pm$	0.9	&	3.64E-03	&	1.23E-04	\\
M31	&	15	&	2CXO J004221.4+411601	&	10.58952 	&	41.26705 	&	13826	&	56084.6894 	&	131.78 	&	0.11	$\pm$	0.07	&	0.0051	$\pm$	0.0004	&	4.48E-02	&	6	$\pm$	3	&	4	$\pm$	1	&	7.32E-03	&	2.68E-04	\\
M31	&	15	&	2CXO J004221.4+411601	&	10.58952 	&	41.26705 	&	13827	&	56090.5510 	&	163.39 	&	0.12	$\pm$	0.06	&	0.0021	$\pm$	0.0003	&	1.72E-02	&	7	$\pm$	4	&	6	$\pm$	2	&	7.32E-03	&	3.32E-04	\\
M31	&	16	&	2CXO J004247.1+411628	&	10.69654 	&	41.27457 	&	11838	&	55343.6374 	&	120.11 	&	0.17	$\pm$	0.08	&	0.009	$\pm$	0.002	&	5.28E-02	&	10	$\pm$	5	&	6	$\pm$	2	&	3.90E-03	&	1.30E-04	\\
M31	&	17	&	2CXO J004212.1+411758	&	10.55074 	&	41.29969 	&	7140	&	54002.8234 	&	536.36 	&	0.06	$\pm$	0.02	&	0.01	$\pm$	0.002	&	1.75E-01	&	0.8	$\pm$	0.3	&	2.6	$\pm$	0.8	&	4.96E-03	&	7.38E-04	\\
M31	&	18	&	2CXO J004303.8+411804	&	10.76611 	&	41.30136 	&	13826	&	56084.9981 	&	737.50 	&	0.07	$\pm$	0.02	&	0.0121	$\pm$	0.0006	&	1.78E-01	&	2.8	$\pm$	0.9	&	8	$\pm$	2	&	3.63E-03	&	7.43E-04	\\
M31	&	19	&	2CXO J004259.6+411919	&	10.74859 	&	41.32203 	&	14195	&	56153.7668 	&	76.56 	&	0.13	$\pm$	0.09	&	0.0042	$\pm$	0.0004	&	3.22E-02	&	40	$\pm$	30	&	16	$\pm$	7	&	3.81E-03	&	8.10E-05	\\
M31	&	20	&	2CXO J004235.2+412005	&	10.64674 	&	41.33495 	&	9524	&	54752.1757 	&	48.45 	&	0.2	$\pm$	0.1	&	0.007	$\pm$	0.001	&	3.63E-02	&	6	$\pm$	4	&	1.4	$\pm$	0.7	&	3.63E-03	&	4.88E-05	\\
M31	&	21	&	2CXO J004210.9+411248	&	10.54567 	&	41.21341 	&	307	&	51572.6431 	&	42.41 	&	0.5	$\pm$	0.2	&	0.0022	$\pm$	0.0009	&	4.67E-03	&	5	$\pm$	2	&	0.6	$\pm$	0.3	&	5.76E-03	&	6.79E-05	\\
M31	&	22	&	2CXO J004310.6+411451	&	10.79423 	&	41.24759 	&	4723	&	53344.4069 	&	259.91 	&	0.21	$\pm$	0.06	&	0.043	$\pm$	0.004	&	2.05E-01	&	5	$\pm$	2	&	5	$\pm$	2	&	4.29E-03	&	3.10E-04	\\
M31	&	23	&	2CXO J004252.5+411854	&	10.71885 	&	41.31513 	&	13828	&	56109.9527 	&	1995.01 	&	0.14	$\pm$	0.02	&	0.046	$\pm$	0.001	&	3.28E-01	&	2.4	$\pm$	0.3	&	21	$\pm$	3	&	3.61E-03	&	2.00E-03	\\
M31	&	24	&	2CXO J004247.0+411910	&	10.69614 	&	41.31973 	&	13827	&	56090.9686 	&	1792.09 	&	0.008	$\pm$	0.005	&	0.0001	$\pm$	0.00005	&	1.19E-02	&	0.13	$\pm$	0.07	&	1.6	$\pm$	0.5	&	9.14E-03	&	4.55E-03	\\
\hline																																					
M81	&	1	&	2CXO J095534.5+690314	&	148.89388 	&	69.05416 	&	18052	&	57419.5764 	&	1766.97 	&	0.017	$\pm$	0.007	&	0.0016	$\pm$	0.0003	&	9.42E-02	&	6	$\pm$	3	&	40	$\pm$	20	&	5.13E-03	&	2.52E-03	\\
M81	&	2	&	2CXO J095510.2+690502	&	148.79279 	&	69.08396 	&	19982	&	57765.7197 	&	209.01 	&	0.1	$\pm$	0.05	&	0.0102	$\pm$	0.0005	&	1.07E-01	&	20	$\pm$	10	&	30	$\pm$	10	&	4.06E-03	&	2.36E-04	\\
M81	&	3	&	2CXO J095500.0+690745	&	148.75025 	&	69.12917 	&	18048	&	57761.8749 	&	8214.26 	&	0.012	$\pm$	0.003	&	0.0058	$\pm$	0.0004	&	5.02E-01	&	4	$\pm$	0.9	&	110	$\pm$	40	&	4.12E-03	&	9.40E-03	\\
M81	&	4	&	2CXO J095524.2+690957	&	148.85121 	&	69.16601 	&	19981	&	57764.6018 	&	127.21 	&	0.16	$\pm$	0.08	&	0.0121	$\pm$	0.0007	&	7.70E-02	&	50	$\pm$	30	&	30	$\pm$	10	&	8.06E-02	&	2.85E-03	\\
M81	&	4	&	2CXO J095524.2+690957	&	148.85121 	&	69.16601 	&	18050	&	57778.1380 	&	329.85 	&	0.11	$\pm$	0.04	&	0.0131	$\pm$	0.0006	&	1.23E-01	&	30	$\pm$	10	&	50	$\pm$	10	&	8.06E-02	&	7.38E-03	\\
M81	&	4	&	2CXO J095524.2+690957	&	148.85121 	&	69.16601 	&	5935	&	53516.1894 	&	94.00 	&	0.16	$\pm$	0.09	&	0.022	$\pm$	0.001	&	1.36E-01	&	20	$\pm$	10	&	16	$\pm$	6	&	8.06E-02	&	2.10E-03	\\
M81	&	4	&	2CXO J095524.2+690957	&	148.85121 	&	69.16601 	&	5936	&	53518.8562 	&	162.25 	&	0.22	$\pm$	0.08	&	0.019	$\pm$	0.002	&	9.04E-02	&	40	$\pm$	20	&	40	$\pm$	10	&	8.06E-02	&	3.63E-03	\\
M81	&	4	&	2CXO J095524.2+690957	&	148.85121 	&	69.16601 	&	5936	&	53518.9470 	&	81.12 	&	0.4	$\pm$	0.2	&	0.019	$\pm$	0.002	&	5.27E-02	&	70	$\pm$	30	&	30	$\pm$	8	&	8.06E-02	&	1.82E-03	\\
M81	&	4	&	2CXO J095524.2+690957	&	148.85121 	&	69.16601 	&	5937	&	53522.4696 	&	217.37 	&	0.14	$\pm$	0.06	&	0.013	$\pm$	0.001	&	9.56E-02	&	40	$\pm$	10	&	30	$\pm$	10	&	8.06E-02	&	4.87E-03	\\
M81	&	4	&	2CXO J095524.2+690957	&	148.85121 	&	69.16601 	&	5938	&	53525.0053 	&	97.24 	&	0.4	$\pm$	0.1	&	0.014	$\pm$	0.001	&	3.97E-02	&	90	$\pm$	30	&	40	$\pm$	10	&	8.06E-02	&	2.18E-03	\\
M81	&	4	&	2CXO J095524.2+690957	&	148.85121 	&	69.16601 	&	5941	&	53533.0073 	&	181.53 	&	0.25	$\pm$	0.08	&	0.017	$\pm$	0.001	&	6.98E-02	&	60	$\pm$	20	&	50	$\pm$	10	&	8.06E-02	&	4.06E-03	\\
M81	&	4	&	2CXO J095524.2+690957	&	148.85121 	&	69.16601 	&	5942	&	53536.0785 	&	131.37 	&	0.3	$\pm$	0.1	&	0.02	$\pm$	0.001	&	7.36E-02	&	50	$\pm$	20	&	29	$\pm$	9	&	8.06E-02	&	2.94E-03	\\
M81	&	4	&	2CXO J095524.2+690957	&	148.85121 	&	69.16601 	&	5943	&	53539.6177 	&	82.66 	&	0.3	$\pm$	0.1	&	0.024	$\pm$	0.001	&	7.97E-02	&	50	$\pm$	20	&	16	$\pm$	6	&	8.06E-02	&	1.85E-03	\\
M81	&	4	&	2CXO J095524.2+690957	&	148.85121 	&	69.16601 	&	5945	&	53545.3099 	&	106.97 	&	0.5	$\pm$	0.1	&	0.026	$\pm$	0.002	&	5.54E-02	&	70	$\pm$	20	&	29	$\pm$	7	&	8.06E-02	&	2.39E-03	\\
M81	&	4	&	2CXO J095524.2+690957	&	148.85121 	&	69.16601 	&	5945	&	53545.3744 	&	240.00 	&	0.17	$\pm$	0.06	&	0.026	$\pm$	0.002	&	1.55E-01	&	25	$\pm$	9	&	28	$\pm$	8	&	8.06E-02	&	5.37E-03	\\
M81	&	4	&	2CXO J095524.2+690957	&	148.85121 	&	69.16601 	&	5948	&	53554.1537 	&	191.24 	&	0.24	$\pm$	0.08	&	0.027	$\pm$	0.002	&	1.16E-01	&	40	$\pm$	10	&	29	$\pm$	8	&	8.06E-02	&	4.28E-03	\\
M81	&	4	&	2CXO J095524.2+690957	&	148.85121 	&	69.16601 	&	5949	&	53557.4171 	&	213.92 	&	0.21	$\pm$	0.07	&	0.03	$\pm$	0.002	&	1.41E-01	&	30	$\pm$	10	&	24	$\pm$	8	&	8.06E-02	&	4.79E-03	\\
M81	&	5	&	2CXO J095617.0+685820	&	149.07088 	&	68.97233 	&	12301	&	55426.9886 	&	3169.68 	&	0.009	$\pm$	0.004	&	0.0018	$\pm$	0.0002	&	1.90E-01	&	2	$\pm$	0.8	&	30	$\pm$	10	&	5.19E-03	&	4.57E-03	\\
\hline																																					
Cen A	&	1	&	2CXO J132526.4-430054	&	201.36008 	&	-43.01509 	&	7797	&	54182.4251 	&	1619.35 	&	0.019	$\pm$	0.008	&	0.0026	$\pm$	0.0002	&	1.40E-01	&	5	$\pm$	2	&	30	$\pm$	10	&	3.33E-03	&	1.50E-03	\\
Cen A	&	2	&	2CXO J132526.7-430125	&	201.36146 	&	-43.02382 	&	7798	&	54186.6903 	&	11619.52 	&	0.008	$\pm$	0.002	&	0.0031	$\pm$	0.0002	&	3.79E-01	&	2	$\pm$	0.5	&	70	$\pm$	30	&	3.49E-03	&	1.13E-02	\\
Cen A	&	3	&	2CXO J132529.6-430122	&	201.37338 	&	-43.02279 	&	22189	&	58597.6691 	&	7688.01 	&	0.007	$\pm$	0.002	&	0.0019	$\pm$	0.0004	&	2.66E-01	&	2.2	$\pm$	0.7	&	80	$\pm$	30	&	3.44E-03	&	7.35E-03	\\
Cen A	&	4	&	2CXO J132525.5-430129	&	201.35613 	&	-43.02495 	&	7800	&	54208.3137 	&	14569.62 	&	0.005	$\pm$	0.001	&	0.0017	$\pm$	0.0002	&	3.54E-01	&	1.2	$\pm$	0.3	&	60	$\pm$	20	&	3.64E-03	&	1.47E-02	\\
Cen A	&	5	&	2CXO J132530.1-430047	&	201.37563 	&	-43.01330 	&	7799	&	54190.0396 	&	458.38 	&	0.03	$\pm$	0.02	&	0.004	$\pm$	0.0002	&	1.22E-01	&	8	$\pm$	5	&	24	$\pm$	9	&	3.33E-03	&	4.24E-04	\\
Cen A	&	6	&	2CXO J132530.8-430041	&	201.37829 	&	-43.01169 	&	7800	&	54207.7463 	&	2560.93 	&	0.021	$\pm$	0.006	&	0.0058	$\pm$	0.0003	&	2.70E-01	&	5	$\pm$	2	&	60	$\pm$	20	&	3.33E-03	&	2.37E-03	\\
Cen A	&	7	&	2CXO J132523.3-430042	&	201.34717 	&	-43.01194 	&	7800	&	54207.9783 	&	1476.35 	&	0.01	$\pm$	0.006	&	0.0006	$\pm$	0.0001	&	5.91E-02	&	2	$\pm$	1	&	20	$\pm$	8	&	3.56E-03	&	1.46E-03	\\
Cen A	&	8	&	2CXO J132532.4-430033	&	201.38513 	&	-43.00928 	&	8490	&	54250.3977 	&	4168.63 	&	0.013	$\pm$	0.004	&	0.0032	$\pm$	0.0002	&	2.43E-01	&	3	$\pm$	1	&	50	$\pm$	20	&	3.35E-03	&	3.87E-03	\\
Cen A	&	9	&	2CXO J132527.4-430214	&	201.36433 	&	-43.03722 	&	11847	&	55455.7761 	&	55.16 	&	0.2	$\pm$	0.1	&	0.006	$\pm$	0.001	&	3.59E-02	&	50	$\pm$	40	&	15	$\pm$	7	&	3.33E-03	&	5.10E-05	\\
Cen A	&	10	&	2CXO J132520.8-430054	&	201.33650 	&	-43.01504 	&	7800	&	54207.8237 	&	172.03 	&	0.06	$\pm$	0.04	&	0.0011	$\pm$	0.0001	&	1.89E-02	&	20	$\pm$	10	&	10	$\pm$	6	&	3.96E-03	&	1.89E-04	\\
Cen A	&	11	&	2CXO J132528.0-430253	&	201.36688 	&	-43.04810 	&	10722	&	55083.1309 	&	1072.05 	&	0.009	$\pm$	0.007	&	0.0003	$\pm$	0.0001	&	3.22E-02	&	2	$\pm$	1	&	9	$\pm$	4	&	3.46E-03	&	1.03E-03	\\
Cen A	&	12	&	2CXO J132535.5-425935	&	201.39788 	&	-42.99308 	&	7800	&	54208.1343 	&	509.25 	&	0.06	$\pm$	0.02	&	0.0028	$\pm$	0.0002	&	4.75E-02	&	14	$\pm$	6	&	26	$\pm$	9	&	7.20E-03	&	1.02E-03	\\
Cen A	&	12	&	2CXO J132535.5-425935	&	201.39788 	&	-42.99308 	&	17890	&	57442.9717 	&	592.19 	&	0.05	$\pm$	0.02	&	0.0024	$\pm$	0.0005	&	4.74E-02	&	14	$\pm$	6	&	60	$\pm$	10	&	7.20E-03	&	1.18E-03	\\
Cen A	&	13	&	2CXO J132517.0-430007	&	201.32096 	&	-43.00205 	&	2978	&	52520.3113 	&	1192.84 	&	0.008	$\pm$	0.006	&	0.0002	$\pm$	0.00009	&	2.39E-02	&	1.2	$\pm$	0.9	&	7	$\pm$	3	&	3.96E-03	&	1.31E-03	\\
Cen A	&	14	&	2CXO J132538.3-430205	&	201.40967 	&	-43.03485 	&	7799	&	54189.8212 	&	19497.11 	&	0.021	$\pm$	0.002	&	0.0136	$\pm$	0.0005	&	6.55E-01	&	5.3	$\pm$	0.6	&	300	$\pm$	70	&	6.66E-03	&	3.61E-02	\\
Cen A	&	14	&	2CXO J132538.3-430205	&	201.40967 	&	-43.03485 	&	7800	&	54207.9369 	&	13812.96 	&	0.018	$\pm$	0.003	&	0.0108	$\pm$	0.0005	&	5.97E-01	&	6	$\pm$	0.8	&	240	$\pm$	70	&	6.66E-03	&	2.56E-02	\\
Cen A	&	15	&	2CXO J132515.6-430210	&	201.31500 	&	-43.03625 	&	8490	&	54250.6265 	&	17627.76 	&	0.0017	$\pm$	0.0007	&	0	$\pm$	0	&	0.00E+00	&	0.6	$\pm$	0.2	&	60	$\pm$	10	&	5.90E-03	&	2.89E-02	\\
Cen A	&	16	&	2CXO J132518.2-430304	&	201.32604 	&	-43.05137 	&	7797	&	54181.6726 	&	1572.17 	&	0.1	$\pm$	0.02	&	0.0471	$\pm$	0.0008	&	4.63E-01	&	27	$\pm$	5	&	130	$\pm$	30	&	4.47E-03	&	1.95E-03	\\
Cen A	&	17	&	2CXO J132541.7-430214	&	201.42379 	&	-43.03738 	&	8490	&	54251.1091 	&	13.15 	&	0.4	$\pm$	0.4	&	0.0001	$\pm$	0.00003	&	2.63E-04	&	100	$\pm$	100	&	12	$\pm$	6	&	4.71E-03	&	1.72E-05	\\
Cen A	&	18	&	2CXO J132511.5-430226	&	201.29813 	&	-43.04066 	&	8490	&	54250.4413 	&	4472.31 	&	0.006	$\pm$	0.002	&	0.0008	$\pm$	0.0001	&	1.43E-01	&	1.7	$\pm$	0.8	&	40	$\pm$	10	&	3.65E-03	&	4.54E-03	\\
Cen A	&	19	&	2CXO J132524.9-430425	&	201.35371 	&	-43.07371 	&	8489	&	54229.4160 	&	14060.48 	&	0.003	$\pm$	0.001	&	0.0005	$\pm$	0.0001	&	1.56E-01	&	0.7	$\pm$	0.2	&	40	$\pm$	10	&	3.65E-03	&	1.43E-02	\\
Cen A	&	20	&	2CXO J132507.3-430407	&	201.28033 	&	-43.06884 	&	7797	&	54182.1159 	&	5267.64 	&	0.014	$\pm$	0.004	&	0.0029	$\pm$	0.0002	&	2.04E-01	&	4	$\pm$	1	&	80	$\pm$	20	&	4.46E-03	&	6.52E-03	\\
Cen A	&	21	&	2CXO J132502.7-430243	&	201.26121 	&	-43.04536 	&	7797	&	54181.7127 	&	10095.75 	&	0.019	$\pm$	0.003	&	0.01	$\pm$	0.0004	&	5.31E-01	&	4.6	$\pm$	0.7	&	140	$\pm$	40	&	8.92E-03	&	2.50E-02	\\
Cen A	&	21	&	2CXO J132502.7-430243	&	201.26121 	&	-43.04536 	&	8489	&	54229.2025 	&	4463.36 	&	0.031	$\pm$	0.006	&	0.0165	$\pm$	0.0005	&	5.26E-01	&	8	$\pm$	1	&	110	$\pm$	30	&	8.92E-03	&	1.11E-02	\\
Cen A	&	22	&	2CXO J132546.5-425702	&	201.44404 	&	-42.95078 	&	10722	&	55083.0452 	&	6255.10 	&	0.01	$\pm$	0.003	&	0.003	$\pm$	0.0003	&	3.13E-01	&	1.7	$\pm$	0.5	&	50	$\pm$	20	&	3.62E-03	&	6.28E-03	\\
Cen A	&	23	&	2CXO J132552.7-430546	&	201.46971 	&	-43.09628 	&	7799	&	54189.5648 	&	188.10 	&	0.13	$\pm$	0.06	&	0.0009	$\pm$	0.0001	&	6.77E-03	&	40	$\pm$	20	&	40	$\pm$	10	&	1.42E-02	&	7.41E-04	\\
Cen A	&	23	&	2CXO J132552.7-430546	&	201.46971 	&	-43.09628 	&	7800	&	54208.1664 	&	138.43 	&	0.18	$\pm$	0.08	&	0.0008	$\pm$	0.0001	&	4.43E-03	&	50	$\pm$	20	&	26	$\pm$	8	&	1.42E-02	&	5.45E-04	\\
Cen A	&	23	&	2CXO J132552.7-430546	&	201.46971 	&	-43.09628 	&	8490	&	54250.5813 	&	286.50 	&	0.12	$\pm$	0.05	&	0.0009	$\pm$	0.0001	&	7.37E-03	&	30	$\pm$	10	&	60	$\pm$	10	&	1.42E-02	&	1.13E-03	\\
Cen A	&	24	&	2CXO J132601.3-430528	&	201.50533 	&	-43.09126 	&	7800	&	54207.9542 	&	26873.43 	&	0.031	$\pm$	0.002	&	0.0197	$\pm$	0.0007	&	6.38E-01	&	8.4	$\pm$	0.7	&	600	$\pm$	100	&	4.76E-03	&	3.55E-02	\\
Cen A	&	25	&	2CXO J132546.0-431103	&	201.44163 	&	-43.18433 	&	8490	&	54250.5871 	&	245.41 	&	0.08	$\pm$	0.04	&	0.0038	$\pm$	0.0003	&	4.66E-02	&	60	$\pm$	30	&	50	$\pm$	20	&	6.63E-03	&	4.52E-04	\\
Cen A	&	26	&	2CXO J132525.2-431448	&	201.35496 	&	-43.24665 	&	8489	&	54229.7072 	&	560.34 	&	0.06	$\pm$	0.02	&	0.0011	$\pm$	0.0001	&	1.76E-02	&	11	$\pm$	4	&	39	$\pm$	8	&	3.12E-02	&	4.86E-03	\\
\hline
\end{longtable}
\end{ThreePartTable}
}
{\fontsize{8.5pt}{12.75pt}\selectfont
\setlength{\tabcolsep}{2.pt}
\renewcommand{\arraystretch}{0.9}
\begin{ThreePartTable}
\begin{TableNotes}[flushleft]
\footnotesize
\item \textit{Notes.—}
(1) Galaxy: host galaxy.
(2) Seq: dip sequence ID; labels correspond to those used in Figures~\ref{fig:acis}.
(3) CXO name: designation in the \textit{Chandra} Source Catalog (v2.1.1). 
(4)–(5) RA/Dec: J2000 coordinates (decimal degrees) of the X-ray source centroid.
(6) ObsID: \textit{Chandra} observation identifier.
(7) Start Time (MJD): start time of the dip block identified by the Bayesian-Blocks algorithm \citep{2013ApJ...764..167S}.
(8) Duration (s): temporal extent of the dip block.
(9) $CR_{\rm dip}$ (cts\,s$^{-1}$): count rate within the dip block.
(10) $CR_{\rm ref}$ (cts\,s$^{-1}$): reference (non-dip) count rate measured from the reference quiescent block.
(11) Depth: fractional decrement relative to the reference block, $\mathrm{depth}=1-CR_{\rm dip}/CR_{\rm ref}$.
(12) $L_{\rm ref}$ ($\rm 10^{38}~erg~s^{-1}$): absorbed 0.5–8\,keV luminosity corresponding to $CR_{\rm ref}$, converted using the instrumental response of the parent observation, the spectral assumptions described in Section~\ref{sec:data}. All the quoted uncertainties are $1\sigma$. 
\item A machine-readable version of this table is available in the online journal.
\end{TableNotes}

\begin{longtable}{ccccccccccccc} 
\caption{Dips in M31, M81, and Centaurus A}\label{tab:dip}\\
\toprule
Galaxy	&	Seq	&	CXO name	&	RA	&	Dec	&	ObsID	&	Start Time	&	Duration	&	$CR_{\rm dip}$	&	$CR_{\rm ref}$	&	depth	&	$L_{\rm ref}$	&	\\
	&		&		&	deg	&	deg	&		&	MJD	&	s	&	cts s$^{-1}$	&	cts s$^{-1}$	&		&	$\rm 10^{38}~erg~s^{-1}$	&	\\
(1) & (2) & (3) & (4) & (5) & (6) & (7) & (8) & (9) & (10) & (11)&(12) \\
\midrule
\endfirsthead

\midrule
\multicolumn{12}{l}{\small \textit{(continued from previous page)}}\\
\midrule
Galaxy	&	Seq	&	CXO name	&	RA	&	Dec	&	ObsID	&	Start Time	&	Duration	&	$CR_{\rm dip}$	&	$CR_{\rm ref}$	&	depth	&	$L_{\rm ref}$	&	\\
	&		&		&	deg	&	deg	&		&	MJD	&	s	&	cts s$^{-1}$	&	cts s$^{-1}$	&		&	$\rm 10^{38}~erg~s^{-1}$	&	\\
(1) & (2) & (3) & (4) & (5) & (6) & (7) & (8) & (9) & (10) & (11)&(12) \\
\midrule
\endhead

\midrule
\multicolumn{12}{r}{\small \textit{(continued on next page)}}\\
\endfoot

\bottomrule
\insertTableNotes
\endlastfoot

M31	&	1	&	 2CXO J004232.0+411314	&	10.63362 	&	41.22069 	&	15267	&	56156.0664 	&	1013.84 	&	0.006	$\pm$	0.002	&	0.032	$\pm$	0.002	&	0.82 	&	0.47	$\pm$	0.03	\\
M31	&	2	&	 2CXO J004248.5+411521	&	10.70219 	&	41.25589 	&	303	&	51464.3062 	&	764.85 	&	0.038	$\pm$	0.007	&	0.095	$\pm$	0.004	&	0.60 	&	1.03	$\pm$	0.04	\\
M31	&	3	&	 2CXO J004242.1+411608	&	10.67572 	&	41.26895 	&	303	&	51464.3053 	&	758.38 	&	0.029	$\pm$	0.006	&	0.084	$\pm$	0.004	&	0.66 	&	0.93	$\pm$	0.04	\\
M31	&	3	&	 2CXO J004242.1+411608	&	10.67572 	&	41.26895 	&	305	&	51523.2630 	&	996.50 	&	0.023	$\pm$	0.005	&	0.057	$\pm$	0.005	&	0.59 	&	0.81	$\pm$	0.07	\\
M31	&	4	&	 2CXO J004247.1+411628	&	10.69654 	&	41.27457 	&	8193	&	54312.1632 	&	542.39 	&	0.004	$\pm$	0.003	&	0.024	$\pm$	0.002	&	0.85 	&	0.87	$\pm$	0.08	\\
M31	&	5	&	 2CXO J004303.8+411804	&	10.76611 	&	41.30136 	&	310	&	51727.9386 	&	1844.27 	&	0.004	$\pm$	0.001	&	0.045	$\pm$	0.004	&	0.92 	&	0.33	$\pm$	0.03	\\
M31	&	6	&	 2CXO J004213.1+411836	&	10.55478 	&	41.31016 	&	303	&	51464.3015 	&	1022.53 	&	0.028	$\pm$	0.005	&	0.069	$\pm$	0.004	&	0.59 	&	0.84	$\pm$	0.04	\\
M31	&	7	&	 2CXO J004310.6+411451	&	10.79423 	&	41.24759 	&	303	&	51464.3056 	&	724.34 	&	0.047	$\pm$	0.008	&	0.101	$\pm$	0.004	&	0.53 	&	1.17	$\pm$	0.05	\\
M31	&	8	&	 2CXO J004238.6+411603	&	10.66078 	&	41.26771 	&	303	&	51464.3051 	&	810.25 	&	0.08	$\pm$	0.01	&	0.157	$\pm$	0.005	&	0.47 	&	1.77	$\pm$	0.06	\\
\hline					
M81	&	1	&	 2CXO J095532.9+690033	&	148.88733 	&	69.00934 	&	18047	&	57574.1807 	&	44837.13 	&	0.089	$\pm$	0.001	&	0.105	$\pm$	0.001	&	0.15 	&	34.6	$\pm$	0.5	\\
\hline																									
Cen A	&	1	&	 2CXO J132523.0-430145	&	201.34608 	&	-43.02932 	&	20585	&	58596.1874 	&	17357.08 	&	0.0013	$\pm$	0.0003	&	0.0032	$\pm$	0.0003	&	0.59 	&	0.87	$\pm$	0.09	\\
Cen A	&	2	&	 2CXO J132527.4-430214	&	201.36433 	&	-43.03722 	&	7798	&	54186.6821 	&	7627.61 	&	0.0053	$\pm$	0.0008	&	0.0108	$\pm$	0.0004	&	0.51 	&	2.38	$\pm$	0.09	\\
Cen A	&	2	&	 2CXO J132527.4-430214	&	201.36433 	&	-43.03722 	&	7799	&	54190.0566 	&	8777.38 	&	0.0056	$\pm$	0.0008	&	0.0109	$\pm$	0.0004	&	0.49 	&	2.42	$\pm$	0.08	\\
Cen A	&	2	&	 2CXO J132527.4-430214	&	201.36433 	&	-43.03722 	&	7800	&	54207.7417 	&	18745.96 	&	0.003	$\pm$	0.0004	&	0.0062	$\pm$	0.0004	&	0.52 	&	1.35	$\pm$	0.08	\\
Cen A	&	2	&	 2CXO J132527.4-430214	&	201.36433 	&	-43.03722 	&	8490	&	54250.6624 	&	15356.68 	&	0.0013	$\pm$	0.0003	&	0.0051	$\pm$	0.0003	&	0.75 	&	1.14	$\pm$	0.07	\\
Cen A	&	2	&	 2CXO J132527.4-430214	&	201.36433 	&	-43.03722 	&	20794	&	58016.2003 	&	6348.67 	&	0.0004	$\pm$	0.0003	&	0.008	$\pm$	0.0004	&	0.95 	&	1.72	$\pm$	0.08	\\
Cen A	&	3	&	 2CXO J132538.3-430205	&	201.40967 	&	-43.03485 	&	8489	&	54229.6375 	&	3773.71 	&	0.005	$\pm$	0.001	&	0.0149	$\pm$	0.0005	&	0.64 	&	3.1	$\pm$	0.09	\\
Cen A	&	4	&	 2CXO J132538.4-430203	&	201.41008 	&	-43.03431 	&	22189	&	58597.7056 	&	3582.44 	&	0.0029	$\pm$	0.0009	&	0.012	$\pm$	0.001	&	0.76 	&	3.2	$\pm$	0.3	\\
Cen A	&	5	&	 2CXO J132540.5-430114	&	201.41892 	&	-43.02078 	&	20794	&	58015.7276 	&	22482.24 	&	0.0004	$\pm$	0.0001	&	0.0027	$\pm$	0.0002	&	0.85 	&	0.6	$\pm$	0.05	\\

\end{longtable}
\end{ThreePartTable}
    
}
\section{Flares in foreground stars: Gaia-based Distance and Main-sequence Classification}
\label{ap:star}
A number of flares are found in the foreground sources in the four {\it Chandra} fields, which are included in the {it Chandra} sample of stellar flares in \cite{2024ApJ...961..130Z}. We provide their basic information, including sky coordinates, ObsID, flare start time and duration, the peak count rate, the count rate of the reference block, and the flare energy and peak luminosity, in Table~\ref{tab:starflare}. Uncertainties on the count rates, luminosities, and energies are quoted at the $1\sigma$ level. 
While some stellar flares can last for more than one blocks, for simplicity, only the block containing the peak count rate is taken into account for the flare duration and energy. We assumed an absorbed thermal plasma model of the form \texttt{tbabs*apec} to convert count rates to fluxes.
The model parameters were fixed at a hydrogen column density of $N_{\rm H}=8\times10^{20}\ {\rm cm^{-2}}$, a plasma temperature of $kT=2~{\rm keV}$, a metal abundance of $Z=0.3Z_\odot$, and zero redshift.
The normalization of the \texttt{apec} component was set to unity when computing the conversion factors, and the model was evaluated over the 0.5–8.0 keV energy band.

For sources with available Gaia DR3 distances from the GSP-Phot module \citep{2023A&A...674A..32B}, we adopted the catalogued values of \texttt{distance\_gspphot}. 
For sources without valid GSP-Phot distances but with reliable parallaxes (signal-to-noise ratio $\varpi / \sigma_{\varpi} \ge 5$), we computed distances by direct inversion of the parallax after applying the recommended Gaia DR3 zero-point correction following \citet{2021A&A...649A...4L}. 
The zero-point correction was evaluated using the \texttt{gaiadr3\_zeropoint}\footnote{\url{https://pypi.org/project/gaiadr3-zeropoint/}} package, which implements the color–magnitude–ecliptic-latitude-dependent model from \citet{2021A&A...649A...4L}. 
When the model could not be evaluated, a constant fallback of $-0.017$~mas was used. 
For sources with low-S/N or negative parallaxes lacking GSP-Phot estimates, no distance was assigned. 

We classified Gaia counterparts as main-sequence (MS) stars using both photometric and astrophysical-parameter criteria. 
For the photometric method, we constructed the Gaia color--magnitude diagram (CMD) based on the BP--RP color and absolute $G$ magnitude ($M_G = G - 5 \log_{10}(d/10~\mathrm{pc})$), where $d$ is the adopted distance. 
Sources located below or near the empirical MS locus, approximated by
$M_G > 4.5 + 3.5 \times [(BP-RP) - 1.0]$ and $\quad 0 < (BP-RP) < 2.5$,
were classified as MS candidates \citep{2018A&A...616A...8A}. 
For the astrophysical method, we used the GSP-Phot parameters (\texttt{logg\_gspphot} and \texttt{radius\_gspphot}) and required $\log g > 4.0$ and $R_\star < 2~R_\odot$, as expected for main-sequence stars at solar-like metallicities \citep{2023A&A...674A..32B}. Sources satisfying either of the two criteria were flagged as main-sequence stars.
All Gaia DR3 data used here were obtained from the ESA Gaia Archive\footnote{\url{https://gea.esac.esa.int/archive/}} using the \texttt{astroquery.gaia} Python interface.

{\fontsize{7pt}{10.5pt}\selectfont
\setlength{\tabcolsep}{2.5pt}
\renewcommand{\arraystretch}{0.9}

\begin{ThreePartTable}
\begin{TableNotes}[flushleft]
\footnotesize
\item \textit{Notes.—}
(1) Galaxy: field in which the foreground star is detected.
(2) CXO name: designation in the \textit{Chandra} Source Catalog (v2.1.1).
(3)–(4) RA/Dec: J2000 coordinates (decimal degrees) of the X-ray source centroid.
(5) ObsID: \textit{Chandra} observation identifier.
(6) Start Time (MJD): start of the Bayesian-Blocks flare segment.
(7) Duration (s): temporal extent of the flare block.
(8) $CR_{\rm peak}$ (cts\,s$^{-1}$): peak count rate within the flare, measured by subdividing the flare block into five equal time bins and taking the maximum bin value.
(9) $CR_{\rm ref}$ (cts\,s$^{-1}$): reference (non-flare) count rate from the reference quiescent block.
(10) Distance (pc): stellar distance (see Section~\ref{ap:star} for the astrometric catalog and distance inference).
(11) $E_{\rm flare}$ ($\rm 10^{35}~erg$): flare energy (fluence) estimated by summing the excess above the reference level.
(12) $L_{\rm peak}$ ($\rm 10^{32}~erg~s^{-1}$): peak luminosity converted from $CR_{\rm peak}$ using the instrument response and spectral assumptions in Section~\ref{ap:star}.
(13) Main Sequence: flag indicating whether the counterpart is consistent with a main-sequence star (according to the photometric/astrometric criteria described in Section~\ref{ap:star}). 
All the quoted uncertainties are $1\sigma$. 
\item A machine-readable version of this table is available in the online journal.
\end{TableNotes}

\begin{longtable}{ccccccccccccc} 
\caption{Flares in Foreground Stars}\label{tab:fg_flares}\\
\toprule
Galaxy	&	CXO name	&	Ra	&	Dec	&	ObsID	&	Start Time	&	Duration	&	$CR_{\rm peak}$	&	$CR_{\rm ref}$	&	Distance	&	$E_{\rm flare}$	&	$L_{\rm peak}$	&	Main Sequence	\\
	&		&	deg	&	deg	&		&	MJD	&	s	&	cts s$^{-1}$	&	cts s$^{-1}$	&	pc	&	$\rm 10^{35}~erg$ 	&	$\rm 10^{32}~erg~s^{-1}$	&		\\	
(1) & (2) & (3) & (4) & (5) & (6) & (7) & (8) & (9) & (10) & (11) & (12) &(13) \\
\midrule
\endfirsthead

\midrule
\multicolumn{13}{l}{\small \textit{(continued from previous page)}}\\
\midrule
Galaxy	&	CXO name	&	Ra	&	Dec	&	ObsID	&	Start Time	&	Duration (s)	&	$CR_{\rm peak}$	&	$CR_{\rm ref}$	&	Distance	&	$E_{\rm flare}$	&	$L_{\rm peak}$	&	Main Sequence	\\
	&		&	deg	&	deg	&		&	MJD	&	s	&	cts s$^{-1}$	&	cts s$^{-1}$	&	pc	&	$\rm 10^{35}~erg$	&	$\rm 10^{32}~erg~s^{-1}$	&		\\		
(1) & (2) & (3) & (4) & (5) & (6) & (7) & (8) & (9) & (10) & (11) & (12) &(13) \\
\midrule
\endhead

\midrule
\multicolumn{13}{r}{\small \textit{(continued on next page)}}\\
\endfoot

\bottomrule
\insertTableNotes
\endlastfoot	
M31	&	 2CXO J004236.6+411350	&	10.65255 	&	41.23064 	&	14198	&	55811.2265 	&	1152.01 	&	0.04	$\pm$	0.01	&	0.0011	$\pm$	0.0002	&	185.72 	&	3.2	$\pm$	0.6	&	4	$\pm$	1	&	TRUE	\\
M81	&	 2CXO J095508.8+685722	&	148.78679 	&	68.95622 	&	12301	&	55426.7721 	&	2218.78 	&	0.018	$\pm$	0.006	&	0.0004	$\pm$	0.00008	&	111.42 	&	1	$\pm$	0.2	&	0.6	$\pm$	0.2	&	TRUE	\\
M81	&	 2CXO J095613.7+685724	&	149.05746 	&	68.95679 	&	19685	&	57608.8974 	&	2066.15 	&	0.01	$\pm$	0.005	&	0.0007	$\pm$	0.0002	&	185.61 	&	1.1	$\pm$	0.3	&	0.9	$\pm$	0.5	&	TRUE	\\
Cen A	&	 2CXO J132522.3-425717	&	201.34292 	&	-42.95475 	&	7797	&	54181.5946 	&	393.72 	&	0.05	$\pm$	0.03	&	0.0019	$\pm$	0.0002	&	163.30 	&	0.8	$\pm$	0.2	&	4	$\pm$	2	&	TRUE	\\
Cen A	&	 2CXO J132522.3-425717	&	201.34292 	&	-42.95475 	&	7797	&	54182.0209 	&	8069.57 	&	0.012	$\pm$	0.003	&	0.0019	$\pm$	0.0002	&	163.30 	&	2.9	$\pm$	0.5	&	0.9	$\pm$	0.2	&	TRUE	\\
Cen A	&	 2CXO J132522.3-425717	&	201.34292 	&	-42.95475 	&	7799	&	54189.5360 	&	9288.64 	&	0.012	$\pm$	0.003	&	0.0028	$\pm$	0.0002	&	163.30 	&	3.6	$\pm$	0.5	&	0.9	$\pm$	0.2	&	TRUE	\\
Cen A	&	 2CXO J132522.3-425717	&	201.34292 	&	-42.95475 	&	8490	&	54250.2622 	&	2641.02 	&	0.028	$\pm$	0.007	&	0.002	$\pm$	0.0002	&	163.30 	&	4.3	$\pm$	0.6	&	2.1	$\pm$	0.5	&	TRUE	\\
Cen A	&	 2CXO J132522.3-425717	&	201.34292 	&	-42.95475 	&	10722	&	55083.3295 	&	2063.27 	&	0.04	$\pm$	0.01	&	0.0039	$\pm$	0.0003	&	163.30 	&	3.1	$\pm$	0.5	&	2.9	$\pm$	0.7	&	TRUE	\\
Cen A	&	 2CXO J132456.1-430259	&	201.23392 	&	-43.04988 	&	20794	&	58016.1886 	&	7908.80 	&	0.007	$\pm$	0.002	&	0.0006	$\pm$	0.00009	&	373.69 	&	12	$\pm$	2	&	2.7	$\pm$	0.8	&	TRUE	\\
Cen A	&	 2CXO J132546.5-430938	&	201.44383 	&	-43.16068 	&	8490	&	54250.7925 	&	9442.08 	&	0.027	$\pm$	0.004	&	0.0024	$\pm$	0.0002	&	1511.81 	&	1290	$\pm$	90	&	170	$\pm$	20	&	FALSE	\\
Cen A	&	 2CXO J132503.2-430924	&	201.26329 	&	-43.15699 	&	8490	&	54250.4885 	&	16837.00 	&	0.003	$\pm$	0.0009	&	0.0004	$\pm$	0.00009	&	506.34 	&	13	$\pm$	3	&	2.1	$\pm$	0.7	&	TRUE	\\
Cen A	&	 2CXO J132500.9-431028	&	201.25396 	&	-43.17472 	&	10722	&	55082.9068 	&	3392.96 	&	0.007	$\pm$	0.003	&	0.0006	$\pm$	0.0001	&	369.90 	&	5	$\pm$	1	&	3	$\pm$	1	&	TRUE	\\
Cen A	&	 2CXO J132625.7-430253	&	201.60717 	&	-43.04828 	&	7799	&	54190.1219 	&	8379.32 	&	0.011	$\pm$	0.003	&	0.0051	$\pm$	0.0002	&	159.13 	&	2.4	$\pm$	0.4	&	0.8	$\pm$	0.2	&	TRUE	\\
Cen A	&	 2CXO J132625.7-430253	&	201.60717 	&	-43.04828 	&	7800	&	54207.7862 	&	5259.39 	&	0.031	$\pm$	0.005	&	0.0046	$\pm$	0.0003	&	159.13 	&	7.3	$\pm$	0.7	&	2.2	$\pm$	0.4	&	TRUE	\\
CDFS	&	 2CXO J033242.0-275702	&	53.17535 	&	-27.95053 	&	2312	&	51891.9293 	&	881.21 	&	0.09	$\pm$	0.02	&	0.0005	$\pm$	0.00009	&	70.91 	&	0.8	$\pm$	0.1	&	1.2	$\pm$	0.3	&	TRUE	\\
CDFS	&	 2CXO J033242.0-275702	&	53.17535 	&	-27.95053 	&	2312	&	51891.6329 	&	2798.47 	&	0.007	$\pm$	0.004	&	0.0005	$\pm$	0.00009	&	70.91 	&	0.2	$\pm$	0.05	&	0.1	$\pm$	0.05	&	TRUE	\\
CDFS	&	 2CXO J033242.0-275702	&	53.17535 	&	-27.95053 	&	8595	&	54393.3927 	&	6597.78 	&	0.009	$\pm$	0.003	&	0.0006	$\pm$	0.00009	&	70.91 	&	0.43	$\pm$	0.08	&	0.13	$\pm$	0.04	&	TRUE	\\
CDFS	&	 2CXO J033242.0-275702	&	53.17535 	&	-27.95053 	&	9575	&	54400.4785 	&	1698.06 	&	0.027	$\pm$	0.009	&	0.0003	$\pm$	0.00006	&	70.91 	&	0.43	$\pm$	0.08	&	0.4	$\pm$	0.1	&	TRUE	\\
CDFS	&	 2CXO J033242.0-275702	&	53.17535 	&	-27.95053 	&	12043	&	55273.9932 	&	2562.10 	&	0.016	$\pm$	0.006	&	0.0005	$\pm$	0.00008	&	70.91 	&	0.22	$\pm$	0.06	&	0.22	$\pm$	0.08	&	TRUE	\\
CDFS	&	 2CXO J033242.0-275702	&	53.17535 	&	-27.95053 	&	12220	&	55365.5791 	&	3636.80 	&	0.01	$\pm$	0.004	&	0.0004	$\pm$	0.0001	&	70.91 	&	0.2	$\pm$	0.05	&	0.13	$\pm$	0.05	&	TRUE	\\
CDFS	&	 2CXO J033242.0-275702	&	53.17535 	&	-27.95053 	&	16177	&	56939.9400 	&	4908.02 	&	0.039	$\pm$	0.006	&	0.0002	$\pm$	0.00005	&	70.91 	&	1.9	$\pm$	0.2	&	0.54	$\pm$	0.09	&	TRUE	\\
CDFS	&	 2CXO J033242.0-275702	&	53.17535 	&	-27.95053 	&	16450	&	56979.8674 	&	2681.37 	&	0.035	$\pm$	0.008	&	0.0007	$\pm$	0.0001	&	70.91 	&	1.1	$\pm$	0.1	&	0.5	$\pm$	0.1	&	TRUE	\\
CDFS	&	 2CXO J033242.0-275702	&	53.17535 	&	-27.95053 	&	16455	&	57322.6980 	&	225.14 	&	0.09	$\pm$	0.04	&	0.0001	$\pm$	0.00004	&	70.91 	&	0.14	$\pm$	0.04	&	1.2	$\pm$	0.6	&	TRUE	\\
CDFS	&	 2CXO J033249.6-275454	&	53.20694 	&	-27.91498 	&	12223	&	55360.4614 	&	5474.49 	&	0.005	$\pm$	0.002	&	0.0003	$\pm$	0.00007	&	4261.38 	&	600	$\pm$	200	&	200	$\pm$	100	&	FALSE	\\
CDFS	&	 2CXO J033249.6-275454	&	53.20694 	&	-27.91498 	&	17677	&	57341.7232 	&	7348.88 	&	0.017	$\pm$	0.003	&	0.0007	$\pm$	0.0001	&	4261.38 	&	5000	$\pm$	500	&	900	$\pm$	200	&	FALSE	\\
CDFS	&	 2CXO J033258.5-275007	&	53.24383 	&	-27.83543 	&	2405	&	51889.6857 	&	9876.08 	&	0.003	$\pm$	0.001	&	0.0001	$\pm$	0.00006	&	340.31 	&	5	$\pm$	1	&	0.8	$\pm$	0.4	&	FALSE	\\
CDFS	&	 2CXO J033258.5-275007	&	53.24383 	&	-27.83543 	&	8594	&	54406.0039 	&	1572.71 	&	0.01	$\pm$	0.006	&	0.0002	$\pm$	0.00004	&	340.31 	&	2.1	$\pm$	0.8	&	3	$\pm$	2	&	FALSE	\\
CDFS	&	 2CXO J033258.5-275007	&	53.24383 	&	-27.83543 	&	8595	&	54393.9140 	&	162.89 	&	0.09	$\pm$	0.05	&	0.0001	$\pm$	0.00003	&	340.31 	&	1.6	$\pm$	0.7	&	30	$\pm$	20	&	FALSE	\\
CDFS	&	 2CXO J033301.8-275009	&	53.25762 	&	-27.83608 	&	2409	&	51897.3960 	&	4187.46 	&	0.004	$\pm$	0.002	&	0.0001	$\pm$	0.00005	&	936.28 	&	25	$\pm$	8	&	9	$\pm$	5	&	TRUE	\\
\hline
\end{longtable}
\end{ThreePartTable}
}
\label{tab:starflare}

\bsp	
\end{document}